\documentclass[twocolumn]{aastex61}
\pdfoutput=1 
\usepackage{amsmath,amstext}
\usepackage[T1]{fontenc}
\usepackage{apjfonts} 
\usepackage[figure,figure*]{hypcap}
\usepackage{natbib}
\usepackage{hyperref}
\usepackage{footmisc}
\usepackage{graphicx}
\usepackage{textcomp}
\usepackage{gensymb}

\shorttitle{The ZTF Observing System}
\shortauthors{Dekany et al.}

\begin{document}

\title{The Zwicky Transient Facility: Observing System}
\author{Richard Dekany}
\affiliation{Caltech Optical Observatories, California Institute
             of Technology, Pasadena, CA 91125, USA}

\author{Roger M. Smith}
\affiliation{Caltech Optical Observatories, California Institute
             of Technology, Pasadena, CA 91125, USA}

\author{Reed Riddle}
\affiliation{Caltech Optical Observatories, California Institute
             of Technology, Pasadena, CA 91125, USA}

\author{Michael Feeney}
\affiliation{Caltech Optical Observatories, California Institute
             of Technology, Pasadena, CA 91125, USA}

\author{Michael Porter}
\affiliation{Caltech Optical Observatories, California Institute
             of Technology, Pasadena, CA 91125, USA}

\author{David Hale}
\affiliation{Caltech Optical Observatories, California Institute
             of Technology, Pasadena, CA 91125, USA}

\author{Jeffry Zolkower}
\affiliation{Caltech Optical Observatories, California Institute
             of Technology, Pasadena, CA 91125, USA}

\author{Justin Belicki}
\affiliation{Caltech Optical Observatories, California Institute
             of Technology, Pasadena, CA 91125, USA}

\author{Stephen Kaye}
\affiliation{Caltech Optical Observatories, California Institute
             of Technology, Pasadena, CA 91125, USA}

\author{John Henning}
\affiliation{Caltech Optical Observatories, California Institute
             of Technology, Pasadena, CA 91125, USA}

\author{Richard Walters}
\affiliation{Caltech Optical Observatories, California Institute
             of Technology, Pasadena, CA 91125, USA}

\author{John Cromer}
\affiliation{Caltech Optical Observatories, California Institute
             of Technology, Pasadena, CA 91125, USA}

\author{Alex Delacroix}
\affiliation{Caltech Optical Observatories, California Institute
             of Technology, Pasadena, CA 91125, USA}

\author{Hector Rodriguez}
\affiliation{Caltech Optical Observatories, California Institute
             of Technology, Pasadena, CA 91125, USA}

\author{Daniel J. Reiley}
\affiliation{Caltech Optical Observatories, California Institute
             of Technology, Pasadena, CA 91125, USA}

\author{Peter Mao}
\affiliation{Caltech Optical Observatories, California Institute
             of Technology, Pasadena, CA 91125, USA}        

\author{David Hover}
\affiliation{Caltech Optical Observatories, California Institute
             of Technology, Pasadena, CA 91125, USA}

\author{Patrick Murphy}
\affiliation{Caltech Optical Observatories, California Institute
             of Technology, Pasadena, CA 91125, USA}

\author{Rick Burruss}
\affiliation{Caltech Optical Observatories, California Institute
             of Technology, Pasadena, CA 91125, USA}

\author{John Baker}
\affiliation{Caltech Optical Observatories, California Institute
             of Technology, Pasadena, CA 91125, USA}

\author{Marek Kowalski}
\affiliation{Deutsches Elekronen-Synchrotron, D 15738 Zeuthen, Germany}

\author{Klaus Reif}
\affiliation{Bonn-Shutter UG, D 53121 Bonn, Germany}

\author{Phillip Mueller}
\affiliation{Bonn-Shutter UG, D 53121 Bonn, Germany}

\author{Eric Bellm}
\affiliation{Department of Astronomy, University of Washington,
             Seattle, WA 98195}

\author{Matthew Graham}
\affiliation{Department of Astronomy, California Institute
             of Technology, Pasadena, CA 91125, USA}

\author{Shrinivas R. Kulkarni}
\affiliation{Department of Astronomy, California Institute
             of Technology, Pasadena, CA 91125, USA}

\email{rgd@astro.caltech.edu}

\begin{abstract}
The Zwicky Transient Facility (ZTF) Observing System (OS) is the data collector for the ZTF project to study astrophysical phenomena in the time domain.  ZTF OS is based upon the 48-inch aperture Schmidt-type design Samuel Oschin Telescope at the Palomar Observatory in Southern California.  It incorporates new telescope aspheric corrector optics, dome and telescope drives, a large-format exposure shutter, a flat-field illumination system, a robotic bandpass filter exchanger, and the key element: a new 47-square-degree, 600~megapixel cryogenic CCD mosaic science camera, along with supporting equipment. The OS collects and delivers digitized survey data to the ZTF Data System (DS).  Here, we describe the ZTF OS design, optical implementation, delivered image quality, detector performance, and robotic survey efficiency. 
\end{abstract}


\section{Introduction}\label{intro}
The Zwicky Transient Facility (ZTF) is a new private/public observing platform for exploring the transient universe at high observing cadence \citep{Bellm_2018}.  Building on the existing Palomar Transient Factory (PTF) \citep{2009PASP..121.1395L} and intermediate-PTF (iPTF) infrastructure at Palomar Mountain, ZTF consists of both an Observing System (OS), described herein, and an integral new Data System (DS) \citep{2019PASP..131a8003M} for data processing, real-time transient alerts, and data archiving.  In this paper, we describe the entire OS required for execution of the ambitious ZTF science program.

The ZTF OS will scan large areas of the available sky several times per night to search for transient optical events using near-real-time reference image subtraction. In addition to acquiring 240-300 two-band images at each location in the Northern Hemisphere per year in execution of a fast-transient survey, the full ZTF data set will be combined to create a deep reference field of the Northern declination cap in support of source selection for the Dark Energy Spectroscopic Instrument (DESI) survey \citep{Dey:18}. 

ZTF prioritizes field of view over depth to bias transient event detection towards targets that are bright enough for follow-up spectroscopy, a fundamental difference and complimentary function to the Large Synoptic Survey Telescope (LSST).  The ZTF OS is designed with 1.01~arcsecond pixel sampling, selected to match the (ZTF-g, ZTF-r, ZTF-i) band image quality of (2.2, 2.0, 2.0) arcsec full-width at half-maximum (FWHM).  For 30~s exposures during dark time, the 5-sigma limiting magnitude of ZTF in (ZTF-g, ZTF-r, ZTF-i) band is (21.1, 20.9, 20.2).  

By reducing overheads from 46~s (in PTF) to 8.66~s, through faster CCD readout and telescope and dome drive upgrades, we have been able to reduce exposure time from 60~s to 30~s, improving nightly frame rate by a factor of 2.7, and increase open-shutter duty cycle from 57\% to 78\%, another factor of 1.3, all while remaining sky noise limited (darkest Palomar sky signal in ZTF-r band >~25~e\textsuperscript{-}/s/pixel).

At 386~mm $\times$ 395~mm corner to corner and 86.7\% fill factor, the ZTF CCD mosaic has 8\% greater field of view than the 14-inch photographic plates used on the same telescope during the two color Palomar Optical all Sky Survey (POSS) from 1950 to 1957 \citep{Minkowski:63:POSS-I}.   ZTF observes the same number (33,660) of square degrees as POSS to similar depth, in just 8 hours per color, revisiting the same coordinates several times per night. Images are relayed in near real time to Caltech{\textquoteright}s Infrared Processing and Analysis Center (IPAC), where they are processed and compared automatically to detect new transients within minutes of the latest observation.  At a slower rate, IPAC processes fully calibrated data and also houses a legacy archive of all ZTF data.

\vfil\penalty-1000\vfilneg
\section{Observing System Overview}\label{overview}
The ZTF Observing System (OS) delivers efficient, high-cadence, wide-field-of-view, multi-band optical imagery to the ZTF Data System for time-domain astrophysics analysis.  It is currently being used to create an all-sky library of reference images, against which subsequent survey observations are compared \citep{Zackay:16} for transient optical events.  Classification of candidate astrophysical transients is performed within the DS using extensive machine learning to discriminate between real and bogus (i.e. artifacts) transient signals \citep{2019PASP..131c8002M}.  Successful development of the OS required major work effort in several different subsystems in order to achieve our transient detection rate goals \citep{Bellm_2018}.  Key among these are modification of the telescope optical design, upgrades to telescope and dome drive systems, construction of a new large cryogenic imaging camera (ZTFC), robotic filter exchanger (FE), flat-field illuminator (FFI), robotic observing software (ROS), and new telescope control system (TCS).  In terms of total work effort approximately 50\% of all effort was required for the camera subsystem, with telescope upgrades following with $\sim$ 25\% of all work effort.

\section{Optical Design}\label{sec:optical_design}
\subsection{Samuel Oschin 48-inch Telescope}\label{ssec:Oschin_telescope}
ZTF OS relies upon the light-collecting power of the 48-inch (1.2-meter) aperture diameter Samuel Oschin Telescope at Palomar Mountain \citep{Harrington:52:P48}.  This Schmidt-type telescope provides an extraordinarily large etendue, delivering some 47~square degrees of field to the ZTF camera system. Originally designed for curved photographic plates, the telescope was used for the seminal POSS-I and POSS-II surveys (\citealt{Minkowski:63:POSS-I}, \citealt{Reid:91:POSS-II}, \citealt{Djorgovski:98:DPOSS}) leading to major discoveries and legacy catalogs (\citealt{1973UGC...C...0000N}, \citealt{1974Obs....94..319V}, \citealt{Abell:58}) used globally for decades.  The use of modern planar CCD sensors, resident within a vacuum cryostat utilizing a thick optical vacuum window, required significant field curvature compensation and modification of the original optical design. 

The 48-inch diameter telescope at Palomar Observatory was begun in 1939, based upon the recently invented wide-field telescope design of Bernhard Schmidt \citep{Schmidt:31}.  The telescope originally consisted of a concave spherical mirror (diameter 1828~mm), an aperture stop located at the center of curvature of this mirror (6124~mm away), and a singlet aspheric (nominally zero power) corrector plate in the plane of the stop (diameter 1244~mm). This design produced a focal plane of excellent image quality approximately mid-way between stop and primary mirror, having a radius of curvature of 3062~mm.  In 1982, in order to improve delivered image quality, the original singlet aspheric plate was replaced with a contact-cemented achromatic doublet fabricated by Grubb-Parsons in the United Kingdom. In 1986 the telescope was dedicated to Samuel Oschin in recognition of the visionary philanthropic support of science and astronomy made by him and his wife, Mrs. Lynda Oschin.

\subsection{ZTF Optical Modifications}\label{ssec:optical_mods}
\begin{figure*}
\begin{center}
\includegraphics[scale=0.45]{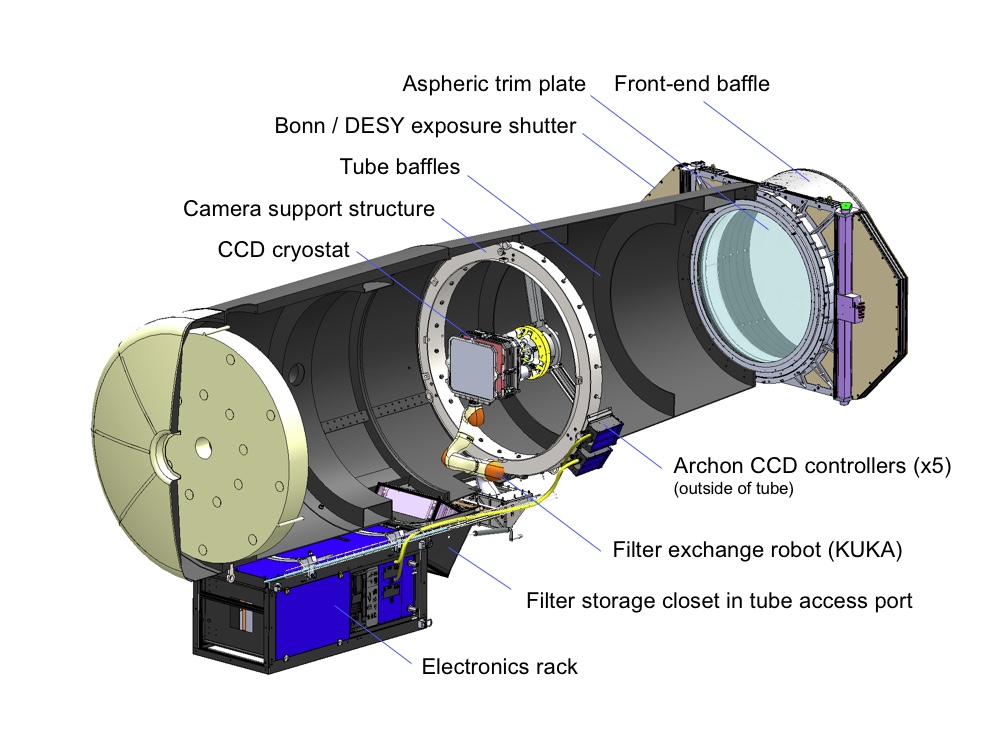}
\caption{Cutaway view of the Samuel Oschin Telescope highlighting new ZTF subsystems.}
\label{fig:p48}
\end{center}
\end{figure*}

To satisfy the ambitious field and image quality requirements for ZTF, several significant optomechanical upgrades to the Samuel Oschin Telescope were required, as shown in Figure~\ref{fig:p48}.  The largest physical change to the optical design is the inclusion of an additional aspheric corrector plate (Section~\ref{sssec:trim_plate}) at the entrance pupil, which corrects for spherical aberration induced by the cryostat window (Section \ref{sssec:cryostat_optics}).  Field aberrations are dominated by spherochromatism and higher-order astigmatism in the science mosaic corners.  Beam walk off of the telescope primary mirror leads to a maximum vignetting of over 30\% relative to the on-axis beam transmission, with most of the sensor area unaffected. 

\begin{figure}
\includegraphics[width=\columnwidth]{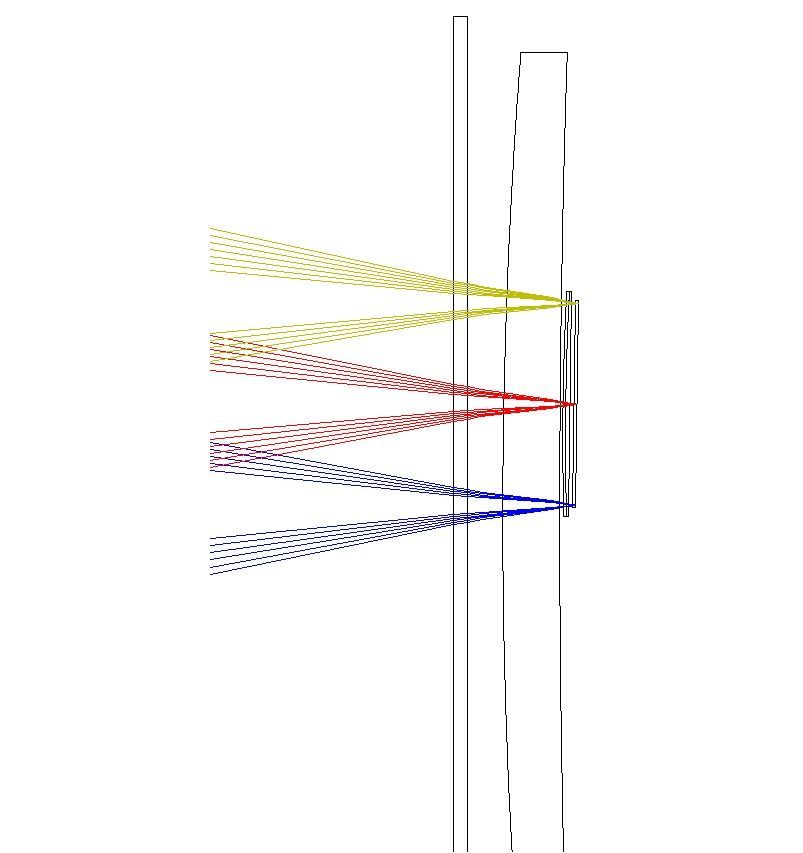}
\caption{ZTF cryostat raytrace showing the spectral filter just outside the dewar, followed by the hierarchical field-flattening solution: the spherical window, individual field-flattening lenses assigned to each CCD, and a spherically tangent CCD mounting plate.}
\label{fig:ztf_zemax}
\end{figure}

A raytrace of astronomical light approaching the cryostat, passing through one of the optical filters, the cryostat window, a single field flattener (Section~\ref{sssec:cryostat_optics}), and onto a single central science CCD (Section \ref{science_CCDs}) is shown in Figure~\ref{fig:ztf_zemax}.  The latter two elements are slightly tilted in this side view.  The distance between filter and convex window vertex is 5~mm, constrained by optical aberrations that arise should the plane parallel filters reside further upstream, and convenient to our filter docking strategy (Section \ref{filter_exchanger}).

The optical design, including individual CCD assembly tip, tilt, and piston on the cold plate was optimized concurrently across the ZTF-r and ZTF-g filters (Section \ref{sssec:filters}) with a merit function designed to maximize the area extent of the field satisfying the delivered image quality (DIQ) error budget (Section \ref{ssec:DIQ}.)  The resultant optical design parameters are summarized in Table~\ref{tab:optics_tab}.

\begin{table*}
\begin{center}
\caption{ZTF Optical Design\label{tab:optics_tab}}
\begin{tabular}{lll}
\hline
\multicolumn{2}{c}{Telescope and Camera Parameters} \\
\hline
Telescope & Palomar 48-\,inch Samuel Oschin Telescope \\
Telescope Type & Prime focus Schmidt with air-spaced triplet corrector\\
Telescope Mount & Equatorial\\
Entrance pupil diameter & 1244.6~mm\\
Working focal ratio & 2.46\\
Camera instantaneous field of view & 46.725~degree$^2$ (ignoring distortion)\\
Plate scale & $1.01^{\prime\prime}$\,pixel$^{-1}$ \\
Filters (Section \ref{sssec:filters}) & ZTF-$g$, ZTF-$r$, ZTF-$i$ \\
Image quality goal (Section \ref{ssec:DIQ}) & $2 - 2.2^{\prime\prime}$\, FWHM in median $1.1^{\prime\prime}$ seeing\\
Optical obscuration & 22.4\% on-axis, up to $\approx 46\%$ at field maximum\\
Design optical distortion, design (max) & 80~arcsec\\
Atmospheric differential distortion & $\approx$ 12~arcsec max (N-S)\\
\hline
\end{tabular}
\end{center}
\end{table*}

\begin{figure}
\begin{center}
\includegraphics[width=\columnwidth]{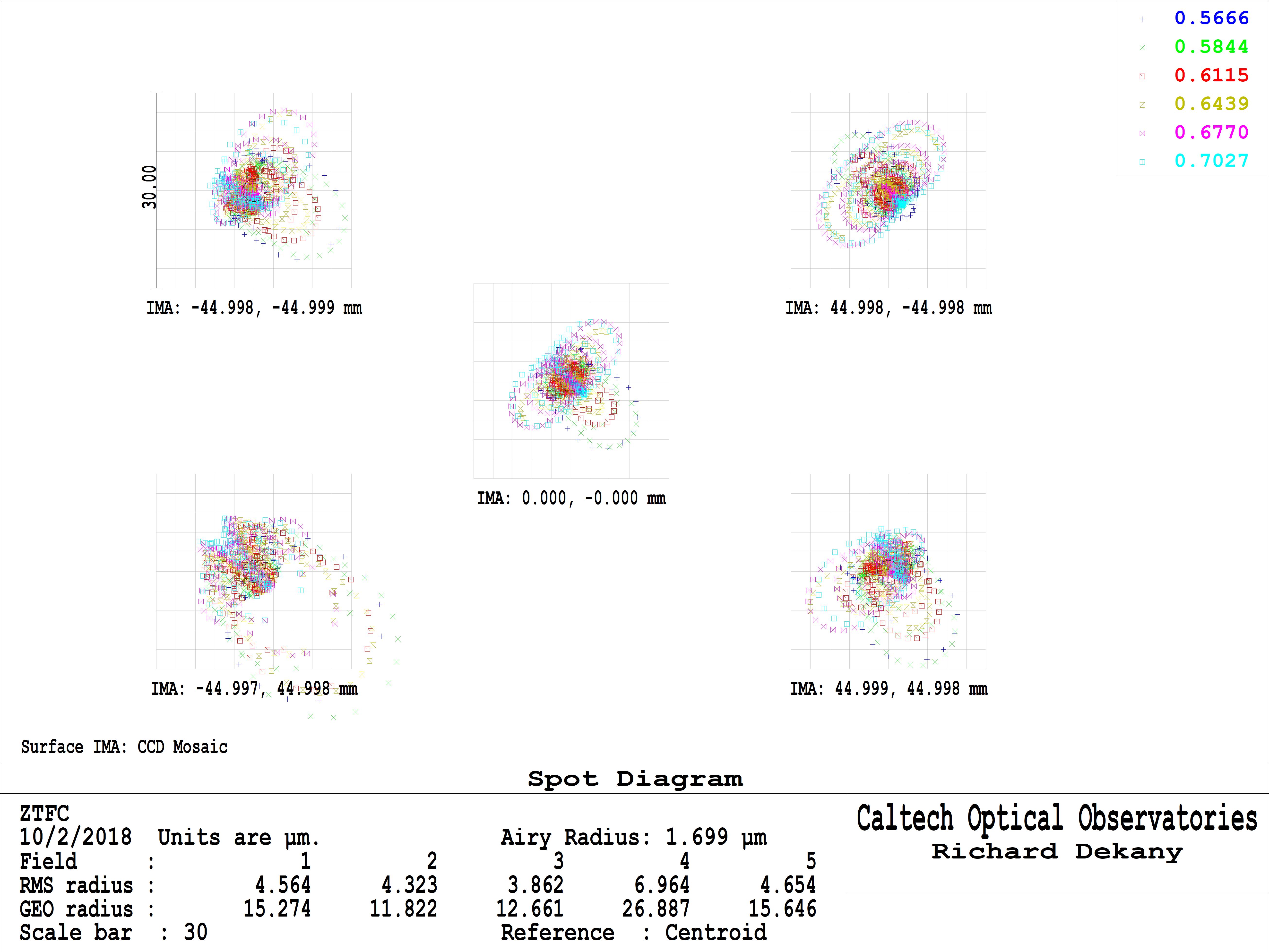}
\caption{ZTF-r band spot diagram for science CCD 'S04', a corner CCD in the 4$\times$4 CCD mosaic.  The lower left spot corresponds to the maximum science field extent of the mosaic (reproduced at 3 other mosaic corners).  The scale bar corresponds to 2 pixels = 2~arcseconds.}
\label{fig:spot_diagram}
\end{center}
\end{figure}

The field-averaged root-mean-squared (RMS) spot sizes, for the optical design aberrations alone, correspond to 0.83 arcseconds and 1.21 arcseconds for the ZTF-r and ZTF-g filters respectively.  A ZTF-r filter spot diagram corresponding to the four corners and center of science CCD 'S04' is shown in Figure~\ref{fig:spot_diagram}.

\subsubsection{Aspheric trim plate}\label{sssec:trim_plate}
The introduction of a cryostat camera into the telescope optical design breaks the original aplanaticity of the Schmidt optical design.  In order to recover sufficient image quality over the ZTF field of view (set to be 2~arcsecond FWHM in order to match Nyquist pixel sampling and maximize field of view), a small change in the aspheric coefficient of the Schmidt corrector was required.  \footnote{In the absence of any telescope prescription change, the inclusion of the cryostat optics alone leads to optical design image quality greater than 4~arcseconds FWHM over the outer 12 of 16 ZTF science CCDs, significantly impacting transient detection rates.}

Several options for compensating the aberrations induced by the cryostat optics were investigated by Caltech and Lawrence Berkeley National Laboratory engineers, including rework of the existing aspheric doublet, fabrication of a new meter-class negative meniscus Maksutov lens, or augmentation of the doublet corrector with a third, air-spaced corrector.  We quickly ruled out any rework of the existing achromatic corrector doublet, fabricated by Grubb-Parsons in the 1980s, due to unacceptable observing downtime and risk, as well as the Maksutov solution due to cost and schedule.  We adopted the strategy of adding a thin, air-spaced aspheric corrector to the existing doublet, at the telescope stop, that would modify the system aspheric coefficient by approximately $-10\%$.  We named this element the 'trim' plate based on its relatively small but important impact to the corrector behavior.  Our strategy was to fabricate a trim plate appropriate for ZTF and install it into the telescope concurrent with the ZTF cryostat installation, as the iPTF survey was continuing to execute active science observations.

\begin{figure}
\includegraphics[width=\columnwidth]{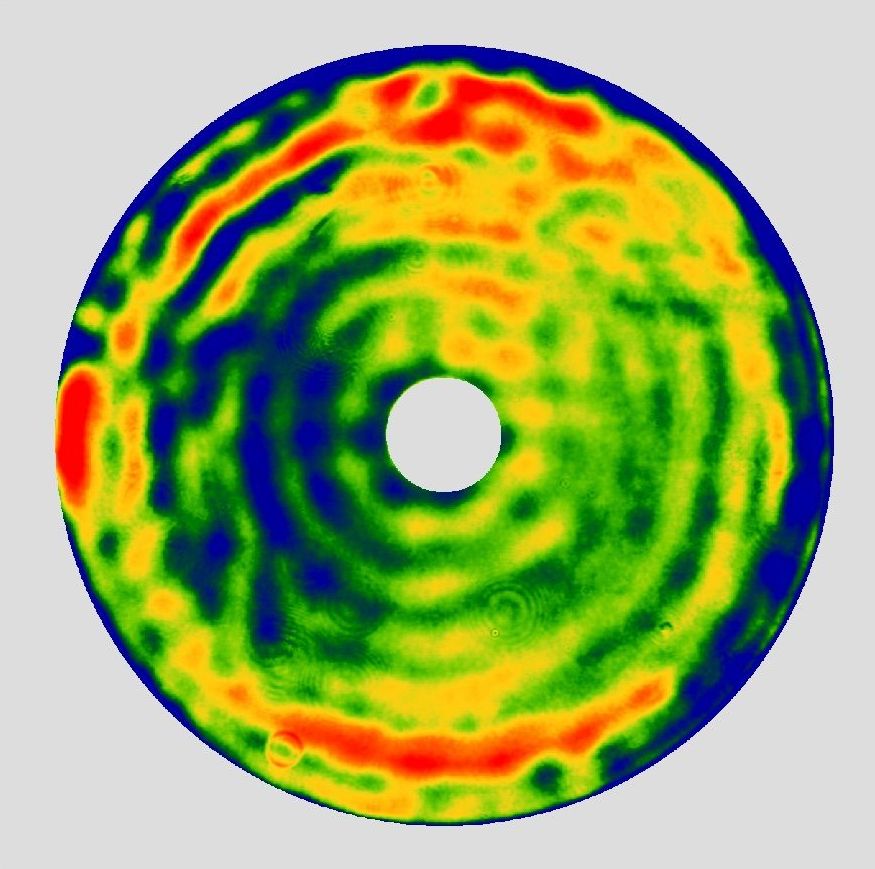}
\caption{Transmission interferogram of the NIAOT trim plate fabricated for ZTF.  The scale is linear with P-V(99\%) = 0.57 waves at 632.8~nm.  The RMS transmitted wavefront error is 0.09 waves at 632.8~nm. (courtesy C. Xu, NIAOT)}
\label{fig:niaot_trim}
\end{figure}

Nanjing Institute for Astronomical and Optical Technologies (NIAOT) in Nanjing, China was chosen to fabricate the trim plate.  A new, 1348~mm-diameter, 15~mm thick, fused silica blank was procured from Corning Advanced Optics of Canton, New York in July 2016 and delivered to NIAOT in November 2016.  NIAOT completed fabrication of this trim plate by August 2017.  The optical figure of the trim plate is shown in the null-interferogram in Figure~\ref{fig:niaot_trim}.  NIAOT delivered the trim plate for two-sided anti-reflection (AR) coating to EMF Corporation of Ithaca, NY, which in turn successfully delivered the plate to Palomar Observatory in November 2017. The mounting of the trim plate is described in Section~\ref{sssec:trim_plate}.

\subsubsection{Spectral filters}\label{sssec:filters}
\begin{figure*}
\includegraphics[width=\columnwidth]{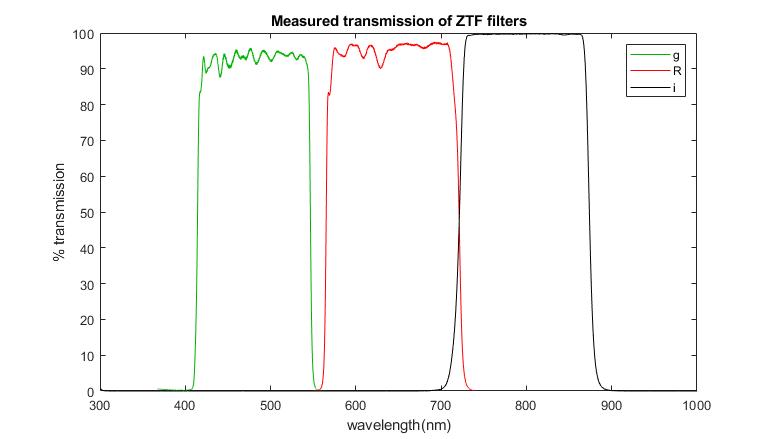}
\caption{Measured transmission of ZTF bandpass filters}
\label{fig:filter_transmission}
\end{figure*}

Filters in the ZTF-g, ZTF-r, and ZTF-i bands are provided to aid classification of  objects.  Each filter is mounted immediately in front of the dewar window (Section~\ref{ssec:filter_assemblies}) and measures 450$\times$490$\times$6.2~mm to cover the 448$\times$415~mm beam footprint. This size complicates manufacture and reduces the number of available vendors. The ZTF-r and ZTF-g filters were produced by Materion (Westford, MA)  The ZTF-i filter was produced by Asahi Spectra (Tochigi, Japan). Measured filter transmission is shown in Figure~\ref{fig:filter_transmission}.  

Development of the filter transmission specifications required several iterations between the science and engineering teams and  detailed discussions with filter vendors to ensure that the filter specifications fulfilled scientific needs while simultaneously being reasonable to fabricate.   The final specification defined four regimes for each filter: a transmission band, a rejection band, and long and short transition bands.  The transition bands were defined to begin the 80\% transmission points, with minimum average transmission between these points required to exceed 93\%.  The location of the transmission bands was specified to within a tolerance of 6~nm.  The rejection band was required to have transmission <~1\%.  The width of the transition bands scaled with wavelength from 15~nm for the green edge of the ZTF-g filter to 31~nm for the red edge of the ZTF-i filter.

\subsubsection{Cryostat Optics}\label{sssec:cryostat_optics}
\begin{figure*}
\begin{center}
\includegraphics[scale=0.38]{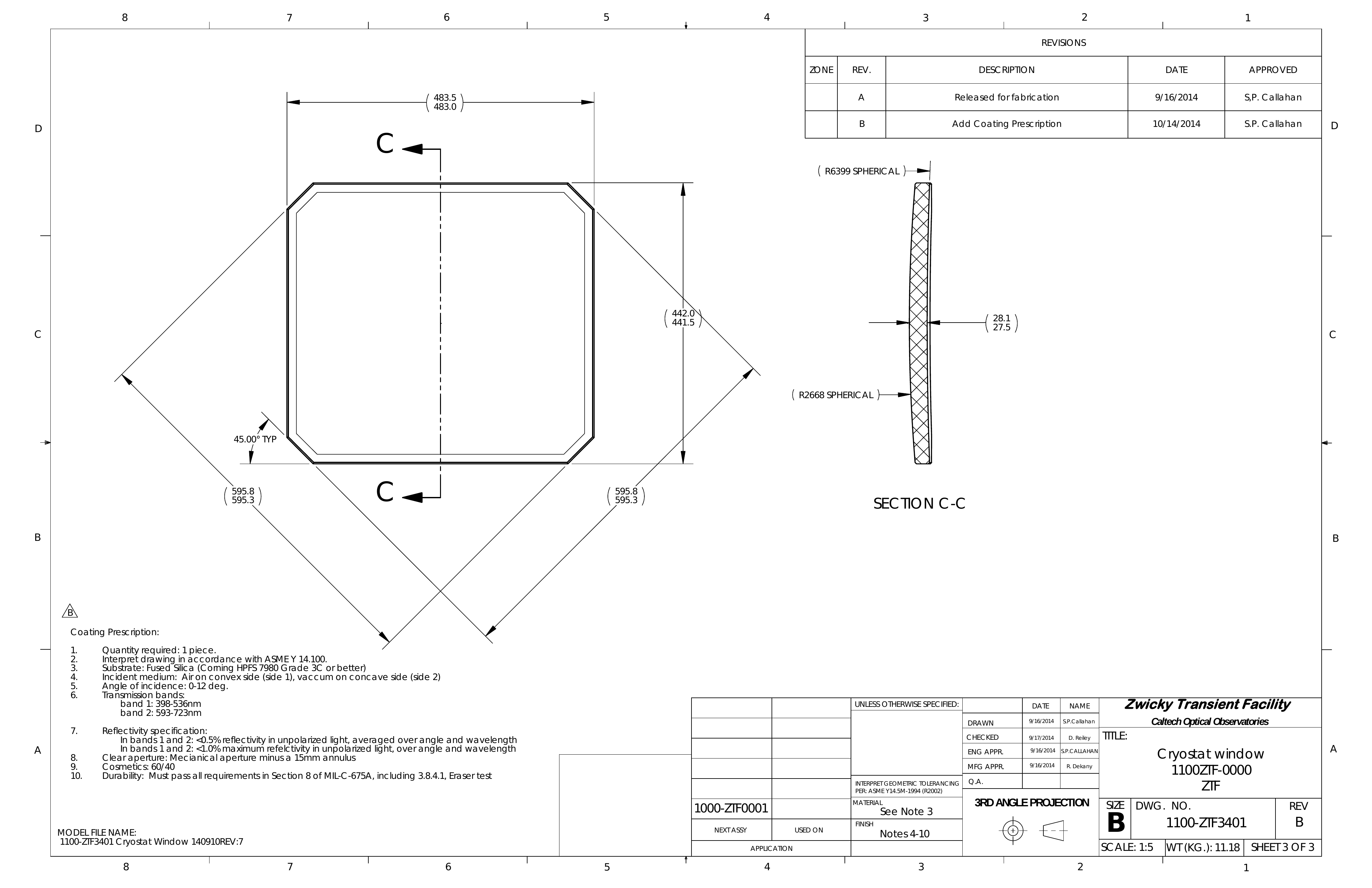}
\caption{ZTF cryostat window as dimensioned for fabrication.}
\label{fig:cryostat_window}
\end{center}
\end{figure*}

After pre-correction by the Schmidt corrector, focusing by the 1.8-meter diameter primary mirror, and spectral filtering, light enters the ZTF cryostat through a spherical, meniscus, fused silica window (Figure~\ref{fig:cryostat_window}).  This window is octagonal in plan view and 442~mm~$\times$~483~mm in size, with 595.5~mm flat-to-flat along the diagonal. The window was figured and polished by Hampton Controls Optics Division of Wendell, PA.  To withstand approximately 2.2~tonnes of atmospheric pressure, the cryostat optical window has 32~mm center thickness\footnote{An initial blank having only 28~mm center thickness was ordered early by the project in an attempt to accelerate schedule, but was abandoned when new analyses suggested prudence for a larger margin of vacuum safety.} and is carefully supported on a series of Viton gaskets to minimize internal window stress during evacuation and over long periods of routine operation.

A conductive indium-tin-oxide (ITO) coating is employed as one of the layers of the anti-reflection coating on the interior surface of the window, to act as an electrical heater that replaces energy lost through black body radiation into the cryostat. Without such heating the center of the window would equilibrate to a temperature well below freezing causing condensation and cold convection cells which would degrade image quality. With 150~$\ohm$/square sheet resistance, about 316~mA at 47~VDC is required to dissipate 15~W typically required to offset the radiative cooling.  Electrical contact is provided by a thin copper foil sandwiched between the window and Viton support gaskets running most of the width of the window at North and South edges.  Power is scaled according to the  Stephan Boltzmann law for radiative transfer using measured tube and CCD temperatures, with scale factor calibrated using a thermal imaging camera in the lab.  Although the window heater has ample capacity and works well in the field, to protect against occasional power losses experienced on the mountain we also  implemented a high capacity dry air supply for the Oschin Telescope that normally provides 4~CFM at sufficiently low humidity to depress dew point to less than $-50\degree$C. In the event of complete power loss, compressed dry air will continue to flow into the tube for 40~minutes and sufficient dew point suppression is sustained for  many hours more.
If the window is made thick enough to fully flatten the focal plane (as was done in PTF), then the central thickness scales with focal plane size. For ZTF the thickness would have been so great that color dependence of focus, due to variation in refractive index across the wide bandpass, would have produced unacceptable loss of image quality. A 32~mm central thickness proved to be  sufficient for mechanical strength, thanks to other optimizations in window support configuration. While 28~mm central thickness would have been  preferred the loss of image quality by increasing to 32~mm could be absorbed in other areas of the image quality budget. The window reduces focal surface curvature from 3062~mm to 7000~mm. 
Residual curvature is accommodated by mounting CCDs on the flat chords of a sphere. 
Finally, to correct for the difference between chords and the sphere, thin plano-convex lenses (convex radius 1260~mm, plano side toward CCD) are mounted 2.0~mm in front of the light-sensitive surface of each CCD.  This leaves slightly less than 1~mm clearance between the frames supporting the field flatteners  and the CCD bond wires.   These field-flattening lenses were produced by Optimax Systems (Ontario, NY).

The frames supporting the field flatteners are thin titanium parts made by direct metal laser sintering (DMLS, 3D printing) which thermally isolate the field flattening lenses so that they serve as ``floating'' radiation shields.  As window, field flatteners and CCDs have unit emissivity at long wavelengths the floating shield halves the thermal load on the cooling system.  The frames are bolted to the sides of the CCD package allowing CCDs to be stored face down facilitating assembly. 

\section{ZTF Mechanical Modifications}
To achieve ZTF observing efficiency goals significant mechanical changes were made to several subsystems of the the Samuel Oschin Telescope in order to both support the wide-field cryostat and reduce observing overheads.

\subsection{Instrument support structure}
\begin{figure*}
\begin{center}
\includegraphics[width=6.5in]{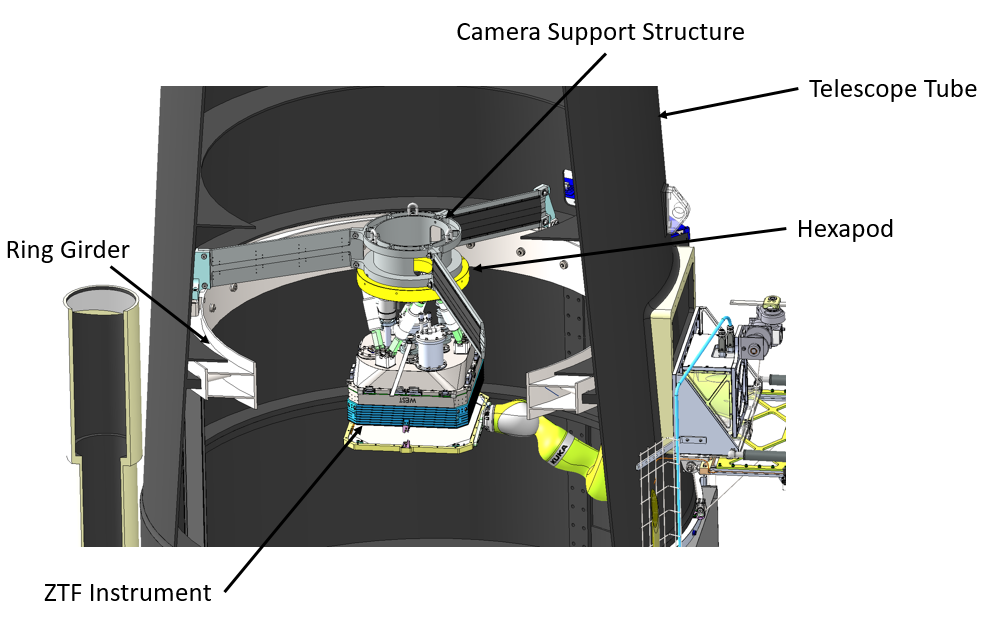}
\caption{Image of the camera support structure with the instrument and hexapod. The structure is comprised of three raked vanes anchored at the top face of the ring girder and tied together at the circular central hub.
\label{fig:CSS_1}}
\end{center}
\end{figure*}

In the original configuration of the Samuel Oschin Telescope, a four-vane structure was used to support a mechanical screw driven focus hub which held the photographic plates at prime focus. The four-vane structure lay in the same plane as the telescope ring-girder and was tensioned to enhance stiffness. This arrangement was adapted for the PTF mosaic, which utilized the same screw driven focus mechanism.  
To accommodate a six degree of freedom hexapod  that provides fine control of tip, tilt, and focus under varying gravity vectors, the entire prime focus assembly had to be replaced. A new cylindrical hub within the instrument footprint is connected by only three vanes to the existing ring girder that provides the interface to the telescope tube. The new vanes are raked backwards to make space for the hexapod.  Unfortunately this  prevents pre-tensioning of the spider vanes, but flexure is small compared to range of the hexapod and considered acceptable given that an autofocus system including tip-tilt sensing was planned (see Figure~\ref{fig:CSS_1}).  The new support structure is mounted on the top face of the ring girder allowing a thicker and thus more rigid interface to be employed since it lies outside the beam. The vanes are beveled to reduce off-axis beam obstruction. To remove play, all  joints are pinned with shoulder screws and orthogonal pre-load set-screws. Electrical cables and refrigerant hoses are located in the hollowed-out center section so that they do not obstruct the beam. This removal of material along the centerline has little impact on vane stiffness. Diffracted power is  reduced by the having less vane area. With 3 vanes diffracted energy is spread among 6 spikes instead of four so that the radial extent of the diffraction pattern is reduced by factor of two, and largely falls within rings formed by ghosting.

\subsection{Telescope and dome drive improvements} \label{ssec:dome_drive}
The previous drive system upgrade performed on the P48 telescope  was implemented by Vertex RSI in 2001 to convert from a multiple motor drive to a single servo motor for each axis.  For ZTF, a new telescope control system was developed to take advantage of modern, commercially available motion control hardware and fully integrated programmable controllers.  The Delta Tau Power PMAC was selected due to its flexibility and adaptability to a wide range of system architectures.  The new servo motors are coupled to the axes via planetary reducers tuned to drive the HA axis at 0.4{\degree}/s\textsuperscript{2}  acceleration and 2.5{\degree}/s maximum velocity and the Dec axis at 0.5{\degree}/s\textsuperscript{2} acceleration and 3{\degree}/s maximum velocity, about twice the speed of the previous drives.  The telescope can reposition through its typical 7{\degree} tiling move, and settle on a new target in 8.8~s in RA and 7.9~s in Dec. 

The dome drive was also upgraded from the original $3/4$~HP motor and reducer which gave 0.2{\degree}/s\textsuperscript{2} acceleration and 3{\degree}/s top speed, to 5~HP motor/reducer driven by a variable frequency drive allowing well-controlled acceleration/deceleration with up to 1{\degree}/s\textsuperscript{2} acceleration, and 5{\degree}/s speed capability.  A rougher paint was applied to the traction path and the single drive tire was also improved to maximize traction and minimize radial pre-load forces on the dome, but currently the acceleration is limited to 0.5{\degree}/s\textsuperscript{2} and the velocity to 3{\degree}/s while assessing the effect on long term reliability.
    
\subsection{Infrastructure, dry air, and lightning protection} \label{ssec:p48_infrastructure}
The P48 observatory electrical power distribution was mostly original 1940s vintage equipment.  To ensure high reliability during the survey, we upgraded the electrical system with modern circuit breakers, switches, and transformers, taking special care to ensure that circuits which powered the sensitive instrument electronics were isolated from possible noise sources such as drive motors.  In addition, we upgraded the entire building and power distribution grounding system to modern standards for sensitive electronic installations, especially for lightning and surge protection.   
To maintain the enclosed telescope tube as a clean and dry environment, we installed an air drying and filtration system.  The dryer is composed of two 5~HP oil-less scroll air compressors, each capable of providing 15~CFM at 100~PSIG and operating in an alternating duplex configuration.  The compressed air is dried by a heat-less, regenerative type dryer, and then filtered by a two-stage filtration system.  The compressed air pressure is regulated and stepped down in pressure before being exhausted into the telescope tube to minimize air's temperature difference with the ambient telescope air.  
An external equipment building was installed adjacent to the P48 dome to house the air compressors, glycol chiller and refrigerant compressors for the Polycold Joule-Thompson cryocoolers.  The combined $\approx 7$~kW of heat produced by this equipment is removed by two HVAC units which work in tandem, or individually if one unit were to fail.     

\subsection{Exposure shutter}
Although a conventional focal plane shutter was used during iPTF, it became clear that any focal plane shutter design, even a rolling curtain, would result in significant ZTF beam obstruction, objectionably reducing the optical transmission.  During the era of photography a shutter consisting of two half-cylindrical panels blocked the light path just below the corrector, stowing against the telescope tube just outside the light path during the exposure. For ZTF we returned to this pupil plane shutter approach, but with a mechanism which opens and closes in just 430~ms to meet observing efficiency goals.

ZTF partner Deutsches Elektronen-Synchrotron (DESY) partnered with Bonn-Shutter UG of Bonn, Germany to design a self contained bi-parting shutter assembly which is attached to the exterior of the telescope tube about 10~cm ahead of the Schmidt corrector assembly.  The specifications of the shutter are shown in Table~\ref{tab:Bonn_Shutter_Spec}.

\begin{table*}
\begin{center}
\caption{ZTF DESY/Bonn-Shutter exposure shutter specifications}
\label{tab:Bonn_Shutter_Spec}
	\begin{tabular}{l l l}
	\hline
	Clear Aperture & 1304~mm \\
	Total mass & $\leq~100$~kg (78~kg delivered)\\	
	Open/close time & $\leq~500$~ms (430~ms delivered)\\
	Open/close time variability & $\leq 10$ ms RMS (2.4~ms RMS, as delivered)\\
	Static load, survival & 200~N over 10~cm $\times$ 10~cm area \\
	Wind load, operability & 15.5~m/s wind, implies 152~N on a single blade \\
	Peak force orthogonal to optical axis & $\leq~3$~N during actuation\\
    Moving mass & 2$\times$3.7~kg per blade $=$ 7.4~kg \\
	Light tightness & $\leq 0.1$ ph/s/pix for a 60~W incandescent bulb located 30~cm\\
	& before the shutter, measured in a $15~\times~15~\mu$ pixel located \\
	&  1.2~m behind the closed shutter ($\leq~0.002$~ph/s/pix delivered)\\
	Dust protection & Hand-vacuum cleaning while closed	
	\end{tabular}
\end{center}
\end{table*}

\begin{figure}
\includegraphics[width=\columnwidth]{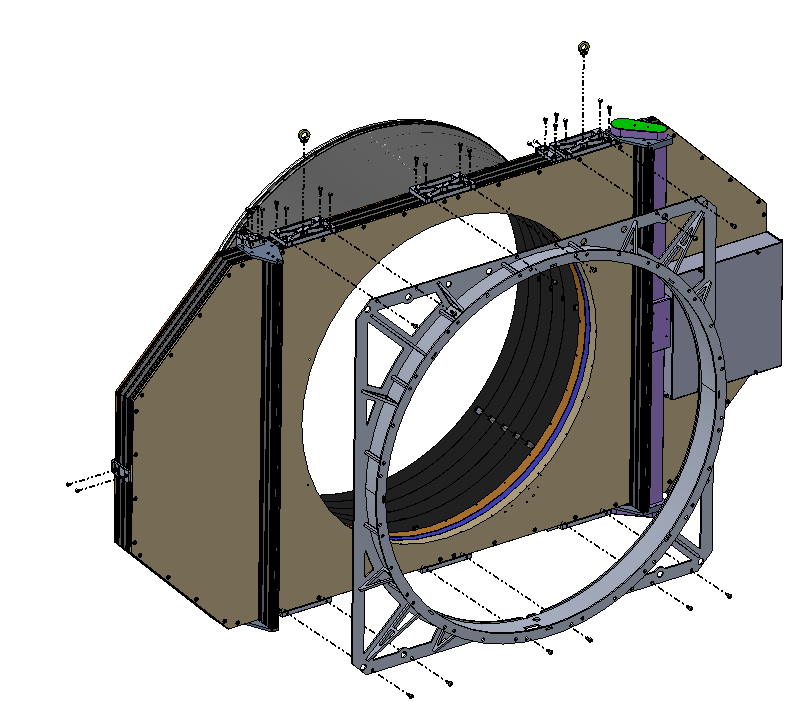}
\caption{The DESY/Bonn-Shutter developed ZTF shutter mounts to the telescope via an Aluminum support structure and supports a new external scattered light baffle.}
\label{fig:Shutter}
\end{figure}

The shutter enclosure is constructed from stock Aluminum extrusions for the border and two cross-bars with Aluminum honeycomb panels for the top and bottom covers.  The shutter blades, constructed from carbon fiber faced honeycomb, each cover 1/2 of the aperture area and do not move independently.  Rather, they move inward or outward symmetrically, coupled to each side of single toothed drive belt and linked to a common servo motor mounted on one of the cross-bars.    The front edge of the blade is a tongue-and-groove arrangement to provide a good light seal during camera read-out.
After extensive testing over a the full range of operating temperatures and gravity vectors at DESY in Zeuthen, the ZTF shutter was delivered to Palomar Mountain in August 2016, where it was employed for a full season of iPTF operation to validate robust winter operations.  Tests conducted on the Oschin Telescope confirmed the peak acceleration of the shutter blades to be 20~m/s\textsuperscript{2}, indicating the imparted force requirement of less than 3~N could be met with blade mass differential of less than 150~grams.  The Rexroth servo electronics allow a wide range of velocity profiles. We adopted a  soft start profile (low jerk), to minimize stress and vibration in the drive train, but have not seen any indication of image quality degradation due to imparted tube vibrations for any of the profiles tested (nor intercepted wind shake) in over two years of shutter operation. This may be due to the very low blade mass imbalance (<~10~grams).  The predominant vibration is acoustic noise at around 1 to 2~kHz, well above telescope resonances.    The shutter is bolted to the telescope front tube flange with a lightweight and stiff aluminum support structure (see Figure~\ref{fig:Shutter}). 

\subsection{Corrector and trim plate lens cells}
\begin{figure}
\includegraphics[width=\columnwidth]{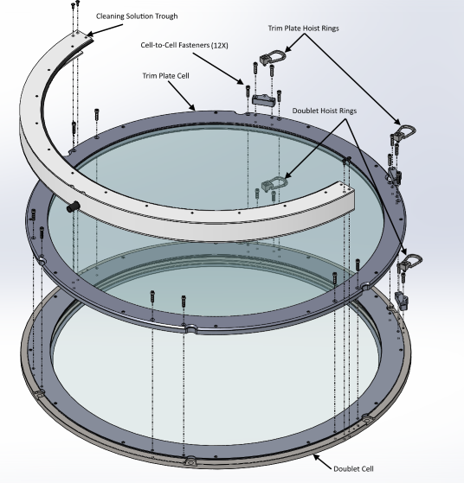}
\caption{Doublet and trim plate cells bolted together.}
\label{fig:DoubletTrim}
\end{figure}

ZTF utilizes both the doublet corrector and trim plate as the fore optics on the front end of the Schmidt telescope (Figure~\ref{fig:DoubletTrim}). Large steel cells were previously used to independently mount either the doublet or the corrector to the front of the telescope, but never both simultaneously. New lightweight aluminum cells were fabricated and addressed some of the technical problems seen with the old cells. The  original doublet corrector used a compliant RTV material to affix the optic in the cell, but over the years the RTV failed and caused the optic to float in an unconstrained manner within the cell. Cork was used for the original trim plate cell and showed similar signs of age.

Making the cells lightweight was critical in keeping the front end mass down and reducing the overall telescope moment of inertia. The mass of the new aluminum cells and optics combined weigh less than just the doublet and its old steel cell.  The doublet and trim plate optics have drastically different thicknesses so the two cells are unique. Despite this, the cells share very similar characteristics. Each optic cell is comprised of two aluminum pieces: 1) a circular cup base and 2) a retaining ring plate. The optics are supported axially on both faces of the glass with a Viton O-ring and radially with adjustable Viton cushions. Additionally, the front face of the trim plate optic is sealed with a double lobed weatherstripping to prevent the ingress of water during the cleaning procedure.  The cell incorporates a gutter and drain to allow the trim plate to be washed with copious quantity of water, while a shield is provide to protect the shutter from getting wet.  This infrastructure is essential since the trim plate is exposed to airborne contaminants including not only dust but pollen and sometimes ash, which are removed with Orvus (a mild soap) and warm water.  A byproduct is that the telescope top end is protected from atmospheric precipitation.

\subsection{Tube baffling}
\begin{figure*}
\begin{center}
\includegraphics[width=6.5in]{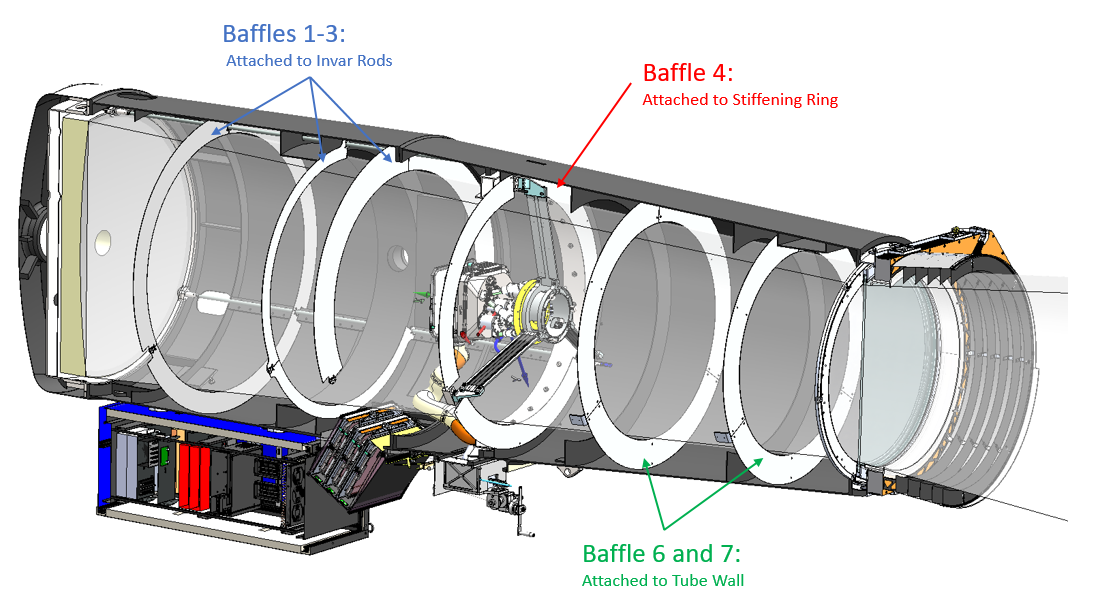}
\caption{ZTF telescope tube baffling layout consists of 7 concentric baffles that off-axis scattered (e.g. moon) light cannot reach the primary mirror without scattering at least twice from blackened surfaces.}
\label{fig:TubeBaffling}
\end{center}
\end{figure*}

 To improve stray light mitigation, the interior of the telescope was lined with lightweight aluminum honeycomb baffles. Each baffle ring is comprised of 3x 120{\degree} sections and attach either to the existing telescope stiffening rings, where available, or to the wall of the telescope with custom machined L-brackets. The baffles were painted with Avian-Black-S paint (2\% reflectance), similar to the interior of the telescope tube, and lined on the undersides with black flocking (<~1\% reflectance). See Figure~\ref{fig:TubeBaffling} for the baffling layout.

\section{Camera}
The key geometric and CCD performance parameters for the ZTF camera are detailed in Table~\ref{tab:ZTF_params}.
\subsection{Cryostat design} \label{ssec:cryostat_design}
The key technical challenge for the ZTF cryostat was to  minimize  obstruction of the relatively small 1.2~m beam by a very large mosaic of science and auxiliary CCDs at the prime focus.  The walls of the vacuum enclosure, the electrical cables, cooling system and associated refrigerant supply hoses have all been within a footprint that is only 9.5~mm larger than the beam footprint on the filter (Figure~\ref{fig:pflow}). The filter has been located only 6~mm from the window to minimize its size, resulting in 22.5\% total on axis beam obstruction, of which 2.3\% is due to the 3 spider vanes. 

\begin{table*}
\begin{center}
\caption{ZTF Camera\label{tab:ZTF_params}}
\label{tab:main_table}
\begin{tabular}{lll}
\hline
\multicolumn{3}{c}{Geometric Parameters} \\
\hline
Plate scale & $1.01^{\prime\prime}$ per 15~$\mu$m pixel \\
Camera ensquared field dimensions & 7.323\degree (E-W) $\times$ 7.504\degree (N-S), measured on sky\\
Light-sensitive area & 1362.5~cm$^2$\\
Inter-detector gaps (active) & 0.205$^\circ$ (E-W), 0.140$^\circ$ (N-S) \\
Focal plane fill factor & 86.7\% \\
Filter exchange time (avg slew) & $\approx$ 80~seconds\\
\hline
\multicolumn{3}{c}{CCD Performance} \\
\hline
& Science ($\times 16$) & Focus ($\times 4$) \\
\hline
CCD model & e2v CCD231-C6 & STA3600A\\
Pixel format & 6144 col $\times$ 6160 row & 2048 col $\times$ 2048 row\\
Pixel width & 15~$\mu$m & 15~$\mu$m\\
CCD package gaps & $\approx$ 0.8~mm & N/A\\
Maximum physical CCD tilt & 3.1$^\circ$ & 3.5$^\circ$\\
Read noise & 10~e\textsuperscript{-} at 1~MHz & 45 to 65~e\textsuperscript{-} at 397~kHz\\
Readout time & 8.2~seconds & 8.2~seconds\\
Gain & 6.2~e\textsuperscript{-}/ADU & 6.6~e\textsuperscript{-}/ADU \\
Pixel full well & 500,000~e\textsuperscript{-} & 220,000~e\textsuperscript{-}\\
Linearity & 1.02\% $\pm$ 0.09\%   (correction factor variation)\\
Linearity saturation & 350,000~e\textsuperscript{-}\\

\hline
\end{tabular}
\end{center}
\end{table*}

\subsubsection{Window support}
In addition to providing the vacuum seal between the stainless steel instrument enclosure and the fused silica window, the O-ring provides the primary mechanical support. The rectangular viton pads provide additional support in the middle of each side, with longer pair compressing a soft copper foil against inner surface of the window to supply a uniform sheet of current to a window heating film deposited on the interior surface. 

The  surface supporting the window was machined to within  $100~\mu$m of the spherical shape of the inner window surface, but deformation of the window shape on this scale would produce fatal stress levels if the glass came in contact with the metal. The durometer and geometry of the supports were carefully optimized to avoid this. The $100~\mu$m groove tolerance prevents excessive stress variations across the window,  unintentional contact with between the surfaces, and vacuum sealing issues.

\begin{figure}
\begin{center}
\includegraphics[width=\columnwidth]{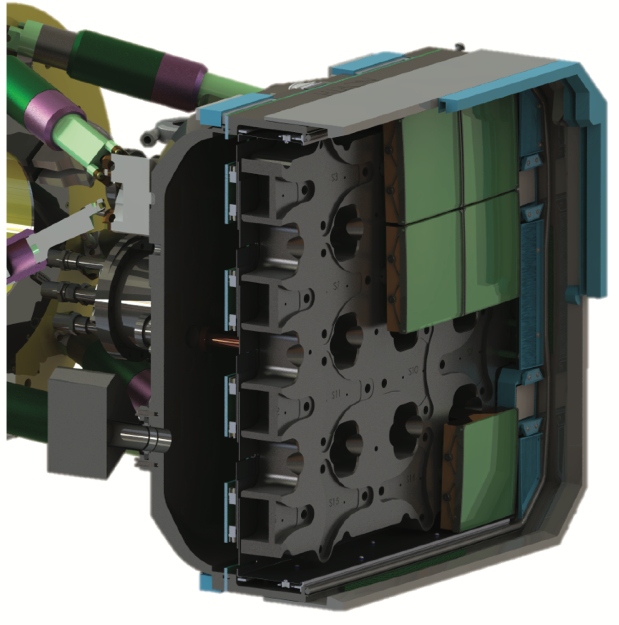}
\caption{This cutaway view of the ZTF camera reveals the faceted cold plate, CCD cable pass-throughs, VIB, and economy of cryostat footprint maximizing optical transmission.}
\label{fig:pflow}
\end{center}
\end{figure}

 The contents of the dewar are installed through the rear so that the supports can be as close to the center of the window as possible without encroaching on the field of view of the CCDs (Figure~\ref{fig:pflow}). The window is surrounded by an Ultem 9085 3D-printed frame which constrains the glass from shifting laterally relative to the rest of the instrument and retains the glass vertically in the event of vacuum loss.

Finite element analysis of the slightly meniscus window predicted a maximum principal stress (at center of the inner surface) of 5.33~MPa.  Using 36~MPa yield stress (1\% probability of failure in 20 years) one calculates a factor of safety of 6.75 in Pasadena, increasing to 8.25 on Palomar. To achieve this figure the deformation along the edges was reduced by locating the supporting pads along the sides but not extending to the corners.  Lower deformation along the edges translates to reduced strain everywhere.  The resulting stress distribution is less centrally concentrated, extending more along the diagonals, resulting in lower peak tensile stress at the center.   Vertical faces at the edge of the window were polished to prevent crack initiation there.  Visual inspection and mild over-pressure tests were employed to test for infant mortality prior to installing any of the ZTF CCDs.

\subsubsection{Field flattener frames}\label{field_flattener_frames}

\begin{figure}
\centering
\includegraphics[width=\columnwidth]{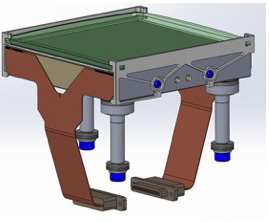}
\caption{A science CCD with its field flattener held between titanium frames, showing signal cables and triad of posts that secure the package against cold plate defining points.}
\label{fig:titframe}
\end{figure}

We trap the 16 field flattener lenses between pairs of thin titanium frames, one attaching to opposing sides of each CCD package as shown in Figure~\ref{fig:titframe}. The frames accurately locate each field flattener above the CCD by referencing the sides of the CCD package, while also providing protection for CCD bond wires. 

The design of the lens pocket allows for thermal contraction differences and manufacturing tolerances, leaving  100~$\mu$m  nominal clearance at operating temperature.  The lens is then constrained axially between two small tabs protruding at each corner with the gap being custom fitted to the corner thickness of each field flattener, for the most part by matching up the dimensional variations of the lenses with those of the frames, and filing some frames where necessary.  Storage trays were fabricated to allow this pairing to be maintained through assembly and disassembly. The plates make lateral contact with the glass at the 45\degree chamfer in the corners and deform elastically to accommodate transverse differential contraction. 

The frames are 3D printed from Ti-6Al-4V Titanium and are black anodized.\footnote{Initially, prototype stainless steel parts were manufactured by GPIprototype using Direct Metal Laser Sintering and scanned in our optical profilometer.  They showed significant but tolerable static deviation from straightness, $\approx200~\mu$m along the 100~mm lens slot.   This level of precision was accommodated in the mechanical tolerancing and the resulting lens displacement proved acceptable optically. The finely stippled surface produced by the DMLS process was black anodized by Techmetal, Inc. using their Techcoat DLA200{\texttrademark} process.}
Previously, we reported on the cryogenic stability of these 3D printed titanium frames.  While we were initially concerned with internal-stress-related warping upon cooling, experiments showed this to be a non-issue \citep{Smith:14:ZTFSPIE}.

Registration with the SiC CCD package is defined by three tabs.   Frames are attached to the CCD with screws which thread into a set of ``nut plates'' embedded into the CCDs.  Flexures incorporated into the nut plate maintain tension on the screw and accommodate thermal contraction, pinned on the side of the registration tab where the fastener is torqued down first and slipping on the other side, where a lower torque is employed.  While the spring action of the nut plate may be sufficient to prevent unscrewing due to thermal walk, a small dot of epoxy between the screw head and the frame is provided as an additional precaution.

The combination of poor thermal contact to the field flatteners and low conductance of the thin titanium steel struts allows the field flattener temperature to be governed by radiative equilibrium so field flatteners function as a floating radiation shield providing almost a factor-of-two cooling power margin with just two Polycold Compact Coolers with PT16 refrigerant. 

There is no AR coating along the two edges of the CCD where silicon has been etched away to expose the front-surface metal for wire bonding. To prevent reflections from both the bond wires and these uncoated surfaces, the nearby edges of the field flattener were inked black (and vacuum baked to eliminate out-gassing). Fortunately the AR coating extends 800~$\mu$m beyond the beam footprint so it is possible to fully obscure the reflective areas without the penumbral zone extending into the image area.  On the other two sides, where the field flatteners enter into a pocket in the frames, the CCD is AR coated all the way to the edges so no masking was applied. Scattering of light from the interior surface of the frames is observed in flat fields suggesting that a narrow mask could have been advantageous. 


\subsection{Thermal control}
Cooling is provided by two independent Polycold Compact Cryocoolers from Brooks Automation. Each Polycold system is comprised of a compressor unit, supplemental eternal charcoal adsorber, 75~ft supply and return lines, and a cold head. The refrigerant blend is a mixture of ethane, methane, propane, and argon and is available in a variety of recipes for different cooling capacity profiles. In the case of ZTF, the PT16 blend is used. Refrigerant charge is adjusted seasonally to maintain supply side pressure at 320 to 350~PSI when operating.  This provides $\sim$~35W total at equilibrium (122-125~K at heat sink), and >~60W peak cooling power at 150~K. At higher temperatures total cooling power drops to 20~W to 30~W resulting in 36 to 48 hour cool down time depending on refrigerant charge pressure and ambient temperature. 

A copper tube counter-flow heat exchanger coil transports refrigerant from a room temperature vacuum flange to the capillary tube at the cold tip in which expanding gas cools the heat sink through gas expansion (the Joule-Thomson effect) or latent heat of vaporization of liquid propane to pre-cool warm incoming gas in the counter-flow heat exchanger. 

Four thermal conductivity swaged copper straps, manufactured by thermal-space.com, link the cold head tips to the instrument's focal plane cold plate. A network of heater resistors evenly distributed on the cold plate heat it to 165~K at center with corners being typically 1~K warmer. The combined effect of contact resistance at the three mounting pads, and $\sim$~1.5W radiative load per CCD allow the CCDs to equilibrate at 170~K. Diode temperature sensors located on the cold heads, thermal straps, and CCDs provide  comprehensive telemetry throughout the instrument. Two Lakeshore 336 temperature controllers are used to servo-control the cold plate heater, refrigerant return line heaters, and backplate heater to ensure that the vacuum enclosure does not cool below ambient.  The telemetry data stream is managed with a Campbell Scientific data logger with a multiplexer expansion.  

To prevent condensation on the instrument window, two copper electrodes make contact with the inner window ITO surface on opposite sides (North-South edges). The copper electrodes span almost the entire width of the window and sit atop a compliant Viton gasket. The inner surface of the window has a thin layer of indium-tin-oxide (ITO) which is optically transparent and electrically conductive. The electrodes are charged with $\pm$48~V and the coating provides a resistance of 220~$\ohm$. A custom temperature controller is used to balance the radiative loads from the telescope tube environment and the internal cold surfaces of the instrument.

Flexible heater ropes spiraling around the refrigerant return lines from the cold heads, through the North camera support structure leg, and out through the exit port on the telescope tube, supply make-up heat to keep the hoses from cooling the spider vanes below ambient.  A byproduct is that condensation no longer forms on the lines when the telescope tube hatch is opened.  Additional temperature sensors are located at the telescope corrector and primary mirror, various locations in the tube and in the dome, and in the equipment shed. 
 
The vacuum interface board (described below) equilibrates to ambient temperature since the radiative losses to the floating radiation shield behind the focal plane fortuitously matches the power dissipation in the preamplifiers, so its temperature is monitored but not actively controlled.

The Archon CCD controllers, mounted on the exterior of the telescope tube near the instrument support structure, are equipped with liquid heat exchangers.  To prevent convection cells from forming, the surface of the electronics rack is maintained within 1~K of ambient temperature through the combination of 2" of insulation in the cabinet walls and chilled water to an internal air heat exchanger equipped with recirculation fans directing cold air across the inside of the cabinet panels before passing across warm equipment. An Optitemp chiller, located in the equipment shed, supplies the telescope with 5~GPM at 40~PSI through 1~inch diameter flexible lines. The chiller set point is automatically adjusted to track ambient dome air temperature minus 2~K so that lines need not be heavily insulated.  A manifold at the electronics rack appropriately distributes the flow between the heat exchanger in the rack and the Archon CCD controllers. 

\section{Image Sensors}
\subsection{Science CCDs}\label{science_CCDs}
In recent years, CCD manufacturers have reduced defect density to the point where lowest cost per unit area is achieved with the largest CCD that fits on a 150~mm silicon wafer.  In 2012 ZTF executed a competitive bid process for the sixteen 6144$\times$6160 pixel CCDs required to completely tile a 47~square-degree instantaneous field of view, selecting e2v, Inc. in February 2013.  The anticipated high yield was confirmed by the delivery of the first batch of six CCD231-C6s within 9 months after placement of the order, with the subsequent 8 CCDs being delivered in the next $\sim$12 months, to match the funding profile.  All CCDs met specifications, and exceeded them in many key areas such as height, flatness, QE, and cosmetic defects. 

Fortuitously, the standard 15~$\mu$m pixel size maps to 1~arcsec providing Nyquist sampling of the 2~arcsec delivered image quality in r band. This image sampling (identical to PTF) provides adequate spatial sampling while minimizing the number of pixels to be digitized, stored and processed.  Furthermore, by sampling the PSF no more finely than Nyquist, signal per pixel is maximized, relaxing read noise requirement so that pixel rate can be higher.  It proved possible to reduce the read time to 8.25~s, safely exceeding the 10~s goal while also meeting or exceeding the 10~e\textsuperscript{-} read noise goal on all channels. 

The only customization of the CCD231-C6 was the AR coating thickness. The first 8 CCDs to be delivered have a single-layer coating optimized for ZTF g and r bands.  QE averaged over all pixels is 82\% at 400~nm rising to 97\% at 500~nm and falling only to 80\% at 700~nm. These devices were placed on the top and bottom two rows of the mosaic.  The CCDs on the central two rows, were purchased in the second batch and employ a two-layer coating with average QE peaking 94\% at 400~nm and staying above 91\% to 650~nm.  

The coarse image scale requires a CCD with the low lateral charge diffusion to minimize PSF degradation.  Since ZTF has no requirement for extended red response, a CCD with conventional thickness and resistivity was selected.  Not only did this improve yield and thus lower the cost, but also has the benefit that lateral charge diffusion is limited to 0.65 pixel FWHM in ZTF-g band and 0.45 pixel in ZTF-r, merely by operating with positively biased image-area clocks to minimize undepleted thickness \citep{Downing:09:CCDs}.

\subsection{Focus Sensors}

For sensing focus, 2k~$\times$~2k CCDs are located near each of the four corners of the science CCD mosaic. Three have an axial offset of $\approx-1.0$~mm (i.e. beyond telescope focus), while the remaining CCD is upstream of focus.  These 100~$\mu$m thick fully depleted n-channel CCDs were designed by Semiconductor Technology Associates, manufactured by Dalsa then delta-doped, AR coated and packaged by the Micro Devices Laboratory at Jet Propulsion Laboratory. The custom package allows for close-butting along two edges of the science focal plane to squeeze the focus CCDs into the optically corrected field and minimize the growth in instrument size, saving a few percent in beam obstruction compared to commercial devices.  The thick, fully depleted silicon and broadband multi-layer coating extends response into the red, while delta-doping extends sensitivity into the UV. 

We utilize the four afocal auxiliary CCDs to make measurements of defocused stars, estimating in each detector respectively a defocus value using the technique of \cite{Tokovinin:06:Donut}.  Using a simple matrix reconstruction, based on a pseudo-inverse of the forward influence matrix, we derive the cryostat hexapod tip, tilt, and focus commands required to maintain a previously established calibrated defocus value vector.

Post-focal images are used to measure cryostat focus, tip, and tilt.  Focus images are processed in real time, directly from memory as soon as they have been read out. All processing is done in C++ and utilizes the open source computer vision library OpenCV.  Out-of-focus donuts are detected and extracted from the image into a collection of postage-stamp images, or regions of interest (ROI). Blends (ROIs which contain multiple, overlapping donuts) are removed from the collection through statistical analysis. A weighting function is applied and the moments are calculated for each donut. A ``donut metric'' is calculated from these moments; since the location of each ROI within the field is preserved in this process,  the donut metric as a function of field position is obtained. Processing time is on the order of a few hundred milliseconds per CCD field, so that the total time to process the images and calculate the correction is approximately 1 to 1.5~s. 

Calculations are completed during a science exposure in preparation for the next science target, with an algorithm that uses a predetermined map for an initial position.  Throughout the night, the donut metric is used to measure and apply updated corrections at each telescope pointing for the focus, tip, and tilt measurements using a simple go-to control loop.  We expect to eventually improve this control law by implementing a modified proportional-integral-derivative (PID) algorithm to reduce the cryostat settling time following large gravity (pointing) change due to telescope slew.

Should image quality gains warrant, given our relatively coarse pixel sampling, room remains for us to pursue more sophisticated analyses. As shown for DESI and LSST, analysis of the intensity distribution across the defocused image will also provide a measurement of higher order wavefront errors that can be diagnostic various misalignment or mirror support problems \citep{Schechter:12:align}. 

Extra-focal-imaging CCDs share the science shutter so exposures are concurrent. The focus CCDs are configured to read from one amplifier per serial register.  Either amplifier may be selected.  

The amplifier at the other end of the serial register was wired as the reference side of a differential signal pair.  Differential readout  would have attenuated the strong clock feed-through enabling faster pixel rate and reducing the line start transient.  However we subsequently realized that the order of the clocks is not the same at both ends of the serial register.  For clock feed-through cancellation to function correctly we should have used the diagonally  opposite amplifier as the reference.

This mistake was unfortunate since these particular CCDs exhibit abnormally large clock feed-through and unexplained slow settling of the video.  These problems were partially mitigated by operation of the Scupper electrode at the highest safe operating voltage (25~V).  Nevertheless we found it necessary to read one focus pixel per three science pixels to allow sufficient settling of transients.   15~e\textsuperscript{-} read noise was achieved at 2.5~$\mu$s/pixel.
 
 Focus CCD Readout is pixel-synchronous with the science CCDs to avoid fixed patterns due to crosstalk.  With 4.5 times fewer pixels per output amplifier, the focus CCDs still read out in less than the science CCDs, so that with sufficiently fast analysis the hexapod can be repositioned before the following exposure.

\subsection{Autoguider}\label{ssec:guide_and_focus_CCDs}

A 4656$\times$3520 pixel monochromatic CMOS sensor is mounted on the bore-sighted guide telescope allowing guiding to proceed even when the science shutter is closed. At 0.49~arcsec per 3.8~$\mu$m pixel, the guider field of view is 38~$\times$~29~arcminutes, roughly centered on the ZTF field.  This field size is ample to guarantee that many sufficiently bright guide stars will be found in all parts of the sky.  

The ZWO ASI1600MM CMOS sensor delivers 60\% QE peaking at 530~nm, and  1.2~e\textsuperscript{-} read noise (at 30~db gain) which is well below photon shot noise from the sky even in the short (1~s) exposures typically used for guiding. A five slot filter wheel (ZWO EFWmini) is populated with standard R, G and B filters and a UV cutoff filter. Focusing is performed via a manual translation stage. 

Custom software was developed to read the guide camera and calculate the guide correction using up to 10 stars in the field, avoiding blended stars or extended features.  Upon command, the initial frame is used to measure the reference centroid positions, with subsequent frames measuring offsets from those positions. The offsets are sent to ROS (which then updates the telescope position) only when the guider software calculates an offset greater than 0.1~arcsec in either RA or Dec.

\section{CCD Readout }
\subsection{Challenges for the electronics}
To minimize the cost of electronics and CCDs, quad readout devices were selected.  For high observational efficiency readout time was required to be <~10~s.  Allowing for  parallel transfer overhead, these requirements lead to a 1~MHz pixel rate on the 64 outputs. Since the back of the instrument is fully occupied by cryocoolers, hexapod, valve, pressure gauge, cables and hoses, the electronics had to be located outside the telescope, presenting significant challenges to achieve the required combination of read noise and pixel rate, while keeping crosstalk low enough to not require any correction in post processing.

The noise requirement was set at 10~e\textsuperscript{-} leading to a 6\% penalty in SNR in 30~sec exposures, given the 25~e\textsuperscript{-}/s minimum dark sky flux in g band.  In r or i bands or with moonlight the SNR penalty is lower.  It was challenging to meet this noise performance at the required pixel rate, given the comparatively long cables and high channel count.

\begin{figure}
\includegraphics[width=\columnwidth]{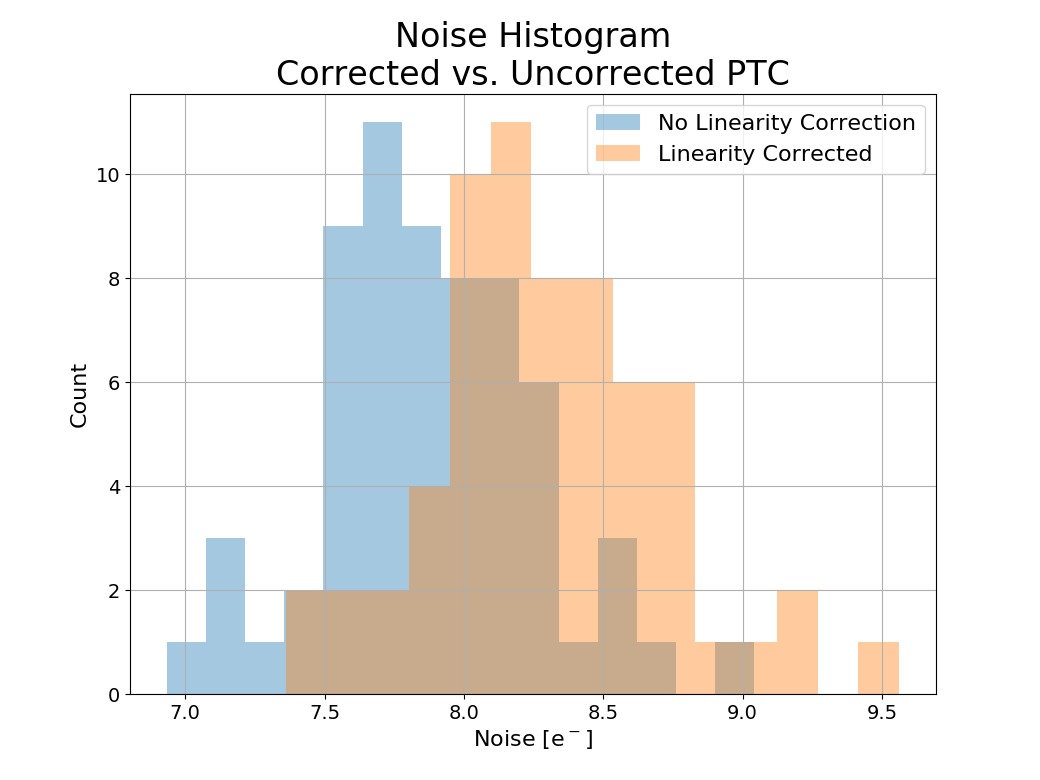}
\caption{Noise histogram for all 64 science channels.}
\label{fig:rn_histogram}
\end{figure}

Design features described below resulted in exceeding requirements by safe margins.  Figure~\ref{fig:rn_histogram} shows the histogram of read noises: most channels exceed the noise requirement in spite of the 8.25~s readout time which also exceeds the requirement.  Cross-talk is undetectable for most channel pairs and less than 1 part in 20,000 on the few where seen. 

\subsection{Vacuum Interface Board}
\begin{figure*}
\begin{center}
\includegraphics[scale=0.25]{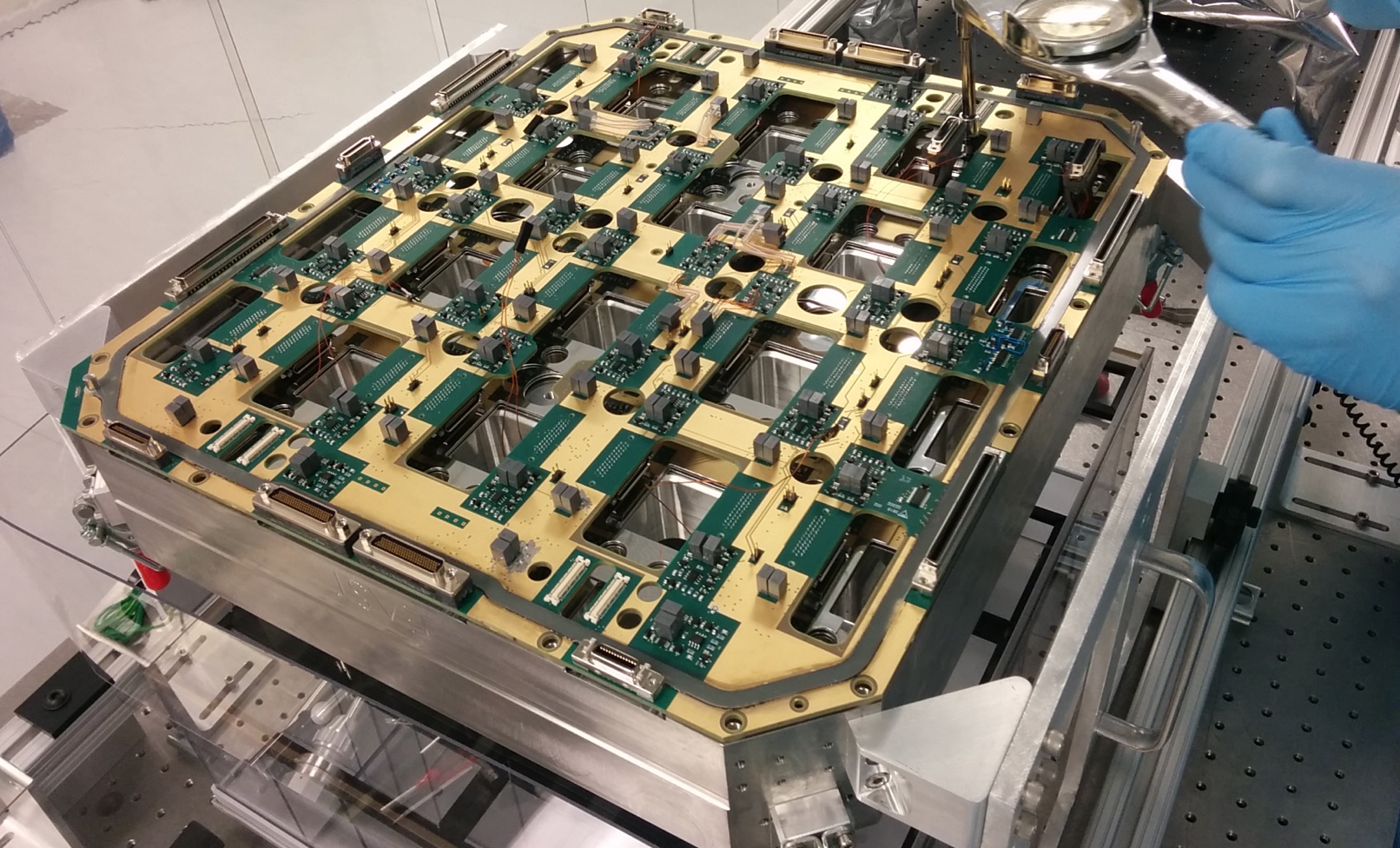}
\caption{The ZTF vacuum interface board (VIB) undergoing cryostat integration at Caltech.}
\label{fig:vib}
\end{center}
\end{figure*}

The Vacuum Interface Board (VIB) is a 1/8" thick, 18 layer printed circuit board trapped between O-rings in side wall and back plate, as shown in Figure~\ref{fig:vib}. This key feature of the ZTF dewar design provides the internal electrical connections and signal preamplification, allowing installation or removal of CCDs in any order, and doing so in a very small footprint. Its primary function is to carry all 1408 electrical connections from 20 CCDs to standard PCB mounted connectors around the periphery, using traces routed on internal layers in place of hermetic connectors.

This packaging method was pioneered in small CCD cameras by \cite{Mackay:13:vib}, and scaled successfully to 14"$\times$22" by \cite{Atwood:13:vib}.  At 19.1"$\times$17.4", ZTF's VIB still fits within the common 24"$\times$18" size limit for many PCB vendors. While bare resin finish has been shown to make a good vacuum seal on other VIBs, we use gold plated copper in ZTF due to its lower emissivity, with the secondary benefit of low outgassing.    

This approach to vacuum feed-through allows connectors to be  chosen for high density and compatibility with off-the-shelf cables.  In ZTF, cables and connectors were made by 3M and Glenair.   In Figure~\ref{fig:vib} one can see that the high density connectors are entirely located within the footprint of the side wall. The cables face rearward and lie close to the tapered rear cover to remain fully hidden within the instrument footprint, as they turn behind the instrument to enter the hollow spider vanes.

Wiring within the dewar is replaced by carefully routed PCB traces arranged for low crosstalk, with controlled characteristic impedance in the case of differential video signals.  Signals are routed around many holes and cut-outs that allow access to CCD fasteners and flex cables, providing the ability for CCDs to be installed or removed in any order.  Two auxiliary signal connectors provide signals for 20 thermal sensors, 5 heaters, high voltage for the window heater, and positive latching signals for the filter which attaches to the cryostat window frame.
 
\subsection{CCD Controllers}

Five Archon CCD controllers, manufactured by Semiconductor Technology Associates (STA) \citep{2014SPIE.9147E..5BB} are mounted outside the telescope tube at the ends of two spider vanes.  They are cooled via conduction with water-cooled heat sinks within each chassis to minimize dome seeing.  There is one controller per mosaic quadrant, plus a 5th controller to operate the four 2k$\times$2k focus CCDs.  Readout must be strictly pixel synchronous to avoid patterns in the images caused by clock-to-video crosstalk, so all controllers share a common master clock and are triggered by a common signal. The four science Archons execute identical code, while the readout for the focus Archon has been padded to remain in synch with the science Archon readouts.

\subsection{Parallel Clocks}
To reduce the number of drivers required, image area clocks are ganged together so that each quadrant appears to be a single 16 channel CCD. The 250~mA clock drive capability of the Archon is substantial compared to many CCD controllers, but the combined capacitive load of four large format CCDs limits  parallel shift time be greater than 275~${\mu}$s.  This would have resulted in 0.85~s contribution to readout time from the 3080 parallel shifts per frame, had we not found a way to hide this overhead completely. We developed a novel clocking scheme 
wherein image area clocks are concurrent with the 2.6~ms line readout. The only delay between lines is the 20~$\mu$s pulse on transfer gate which allows charge to spill into the serial register which is more positively biased than the nearby image area clock electrode. Clock drive current is minimized by making clock waveforms triangular.  Substrate return current is nulled by using the slope control feature of the Archon to tune slopes to compensate for different capacitances to substrate, such that the current into the rising clock is matched to the current drawn from the concurrently falling clock.  The residual fixed pattern is <~10~e\textsuperscript{-} and entirely removed by the standard bias frame subtraction. 

\subsection{Serial and Pixel Clocks}
Each serial clock driver is connected to the same pin on a pair of CCDs.  Given the much lower capacitance of serial clocks, worst-case transition time is 44~ns, but the combination of electrode capacitance and the inductance of the 2~m cable to the Archons results in substantial ringing at the fastest edge rates. To avoid this we invoke the programmable slope control provided by the Archon controller to produce \textit{triangular} clock edges such that there is always one clock falling and another rising while the third is stationary.  Slopes are adjusted to  null the current in the long return path from CCD substrate to controller. Minimizing the slew rate also improves the common mode rejection by the differential signal path, and thus the rejection of residual clock feed-through.

To maximize the fraction of the pixel time during which noise averaging can occur, the reset and charge dump times had to be kept very short.  The 80~ns reset pulse would have been difficult to propagate down the 2~m cable from the Archon without excessive ringing, so it is generated adjacent at the CCD connector (on the VIB) by an EL7457 pin driver, switching between two bias voltages, with current supplied by local capacitors.  The timing pulse is transmitted by Low Voltage Differential Signalling (LVDS) to minimize crosstalk to the analog signals.  

Unfortunately, the same method was not employed for Summing Well, so some additional overhead is incurred by limiting its slew rate to minimize overshoot. This overhead is partially hidden by time allowed for settling of preamplifier after the charge dump edge.  This must be sufficient to avoid non-linearity due to dependence of settling time on signal amplitude.  

\subsection{Video Path}
We utilize the reference output of the e2v CCD231-C6 so that the  signal path is fully differential from the CCD to the AD converter, providing of order hundred-fold suppression of crosstalk and clock feedthrough.  This comes at the cost of  increased theoretical noise by a factor of $\sqrt{2}$.  The reference output is supplied with the same clock and bias signals so that the differential receiver rejects both clock feedthroughs and variation in voltage due to electrical noise or thermal drift.   The suppression of transients has the effect of reducing settling time and thus supports faster pixel rate.

Due to the long path to the controllers, all 72 video signals are amplified on the VIB.  Differential gain is 1.3 while common mode gain is unity.   The preamplifier also serves as a differential line driver for the Mini D Ribbon twinaxial cable from 3M (P/N 141A0-SZ5B-300-0DC), employing 100~$\ohm$ series termination for impedance matching at the driver leaving high impedance at the receiver.   The video signal rides on a DC offset of order 20~V which is removed by 100~nF polypropylene dielectric capacitors on both the signal and reference sides, with analog switches to ground that are asserted once per line to re-establish the offset at the preamplifier input.  

\begin{figure}
\includegraphics[width=\columnwidth]{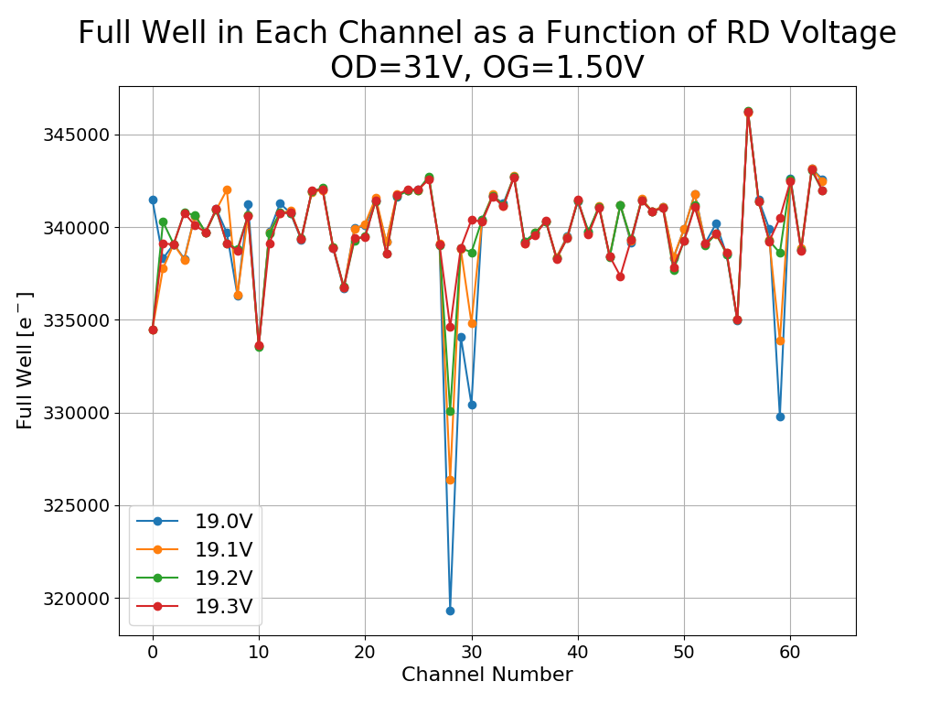}
\caption{Measured ZTF OS detector well capacity.}
\label{fig:well_capacity}
\end{figure}

\subsection{Dynamic Range and Linearity}
Well capacity is  high in the image area thanks to the incorporation of four-phase clocks by e2v.   This is a highly desirable feature since the wide field of view and broad spectral bandpass produces significant number of bright stars in every CCD quadrant, notwithstanding the modest telescope aperture.   By increasing the voltage swing in the serial register and careful optimization of sense node voltage (Reset Drain), Output Gate and Summing Well, we have been able to preserve this well capacity throughout the signal path, consistent the 350,000~e\textsuperscript{-} value promised in the data sheet (see Figure~\ref{fig:well_capacity}).   

\begin{figure}
\includegraphics[width=\columnwidth]{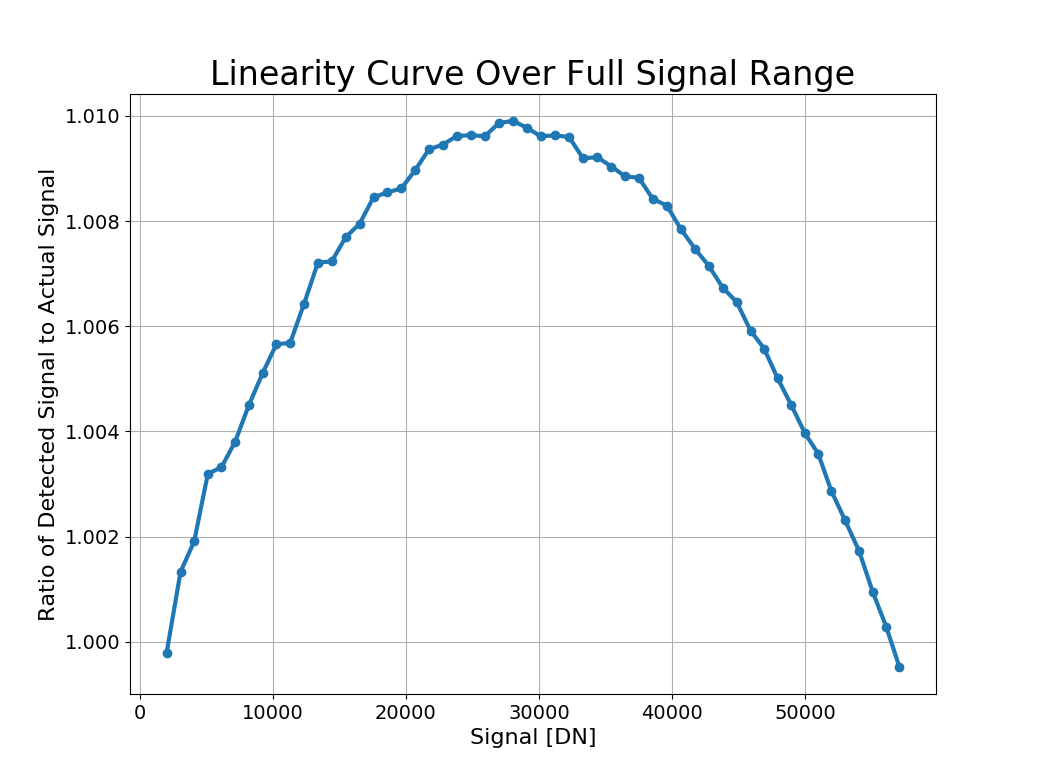}
\caption{Example gain correction factor over full signal range for one channel. 64 such plots were produced, one for each amplifier.}
\label{fig:lin_curve}
\end{figure}

The 7~$\mu$V/e\textsuperscript{-} output node sensitivity converts the $\sim$340~ke\textsuperscript{-} well capacity to a 2.38~V signal swing which produces a gain variation over the signal range that we minimize by adjusting bias voltages, principally Reset Drain.  Linearity optimization would have been very time consuming if using conventional methods (signal versus exposure time for constant flux), so we developed a much faster and more robust method which compared just two exposures having identical illumination.   One is binned on chip with bin factor increasing by one after each line is read out.  This is compared with the equivalent binning performed digitally.  When gain variation over the full signal range is minimized, the transfer function is S-shaped, with peak slope at mid signal levels, and the gain correction factor is well approximated by a quadratic (Figure~\ref{fig:lin_curve}).

\subsection{Digital Correlated Double Sampling}
The dynamic range of the CCD exceeds that of a standard 16-bit AD converter if gain is selected to correctly sample the noise after analog averaging.  Instead, the video is digitized at 100~MHz and with $\sim$~30~MHz analog bandwidth to minimize aliasing of higher frequency noise.  The wider  bandwidth passes more noise to the ADC so that gain can be decreased to bring the signal on scale while still sampling with sufficient resolution to avoid quantization.  The subset of samples within each pixel are designated as baseline or signal.  Each is summed and the difference taken to implement Digital Correlated Double Sampling (CDS).   

The remaining step is to normalize and select which bits to preserve.    We format the result as a 32-bit floating number in which many of the low order bits contain nothing but noise.    These excess noise bits  are discarded as part of the data compression step, based on the measured noise histogram of the data.  Image area and overscan are stored in separate file extensions so that the noise estimation prior to compression correctly estimates the background level which dictates the true noise floor.   The choice of how many bits to preserve prior to compression is analogous to the selection of gain prior to AD conversion in a conventional analog CDS processor.  Advantages of the digital CDS processor include greater dynamic range (as noted), superior rejection of low frequencies since baseline and signal take the same path to the AD converter, and there is no need to adjust electronic gain or time constant when changing the sample integration time or pixel rate.

\subsection{Digital Oscilloscope Mode}
An extremely useful byproduct of digital CDS is that data can be recorded in memory at the full 100~MHz sampling rate without CDS processing.  The Archon controllers support this for one channel at a time, allowing the video waveform to be plotted in near real time.  We made extensive use of this feature to optimize readout timing, noise performance and linearity by remote access during the daytime while ZTF remained mounted on the telescope for commissioning activities at night.   We  close the AC coupler's clamping switch on the reference side to convert the signal path to single sided allowing the clock feedthrough to become visible.  By clamping both sides the shorted-input noise can be measured.

\section{Filter Exchanger}\label{filter_exchanger}

\begin{figure*}
\begin{center}
\includegraphics[scale=0.5]{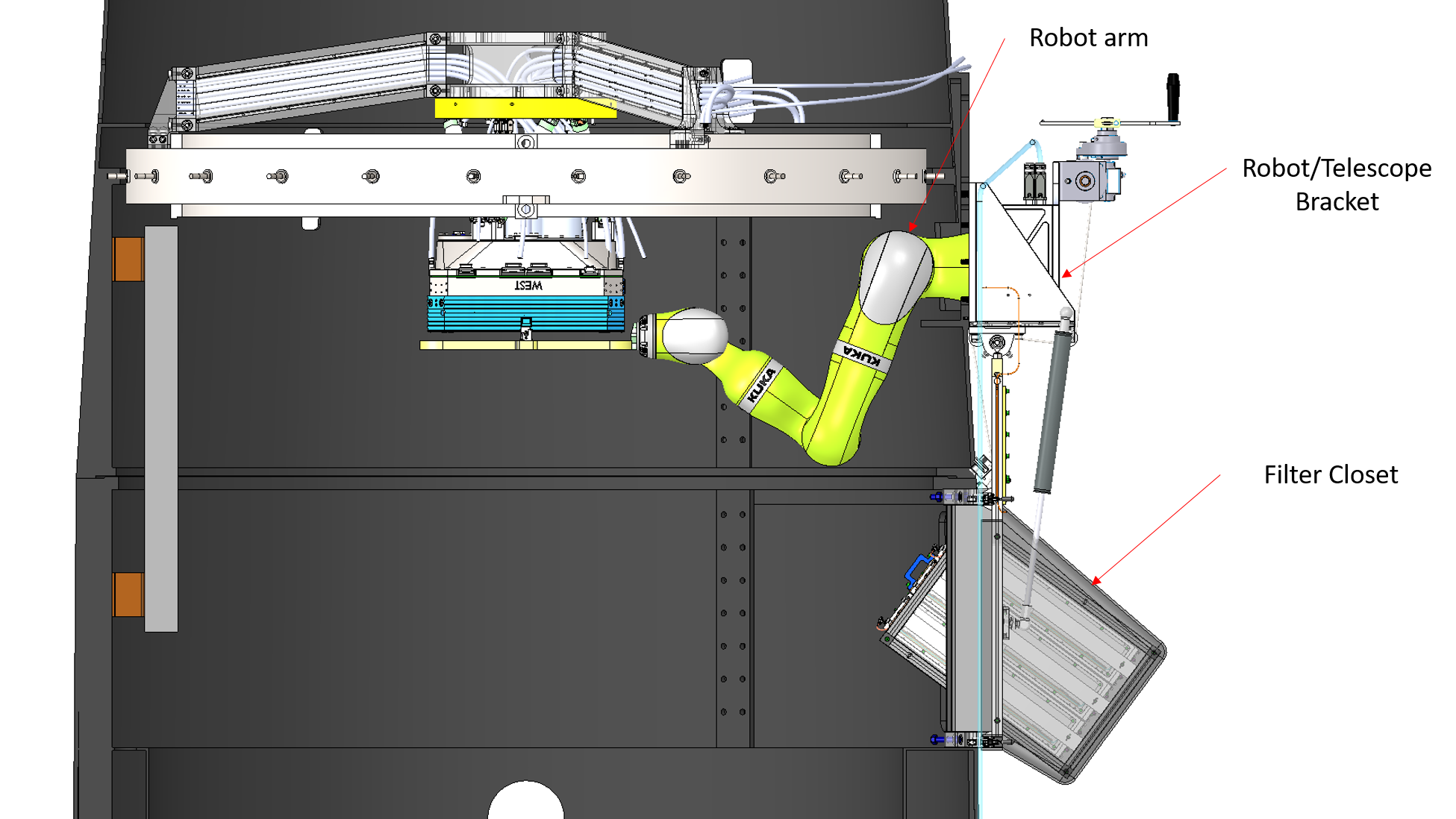}
\caption{Overview of the ZTF Filter Exchanger System.}
\label{fig:fe_crosssection}
\end{center}
\end{figure*}

The ZTF science survey planning set the requirement that optical bandpass filters be exchanged up to 20 times per night (with 90 seconds between exposures to minimize loss of observing time), in order to support both multi-spectral transient classification and band-specific science investigations.  Several architectures of filter exchanger mechanism were considered, including self-contained options (e.g. use of a tank-tread filter turret), a folding petal apparatus, and heritage designs based on the original Schmidt Telescope photographic plate loader which would utilize a long arm and/or permanently affixed rail system.  However, none of these alternatives met project objectives of maintaining high optical transmission to our prime focus camera because they required additional obscuration of the science beam.  Ultimately, an industrial robotic arm was selected as the mechanism to exchange the full-field spectral filters for ZTF.  

The robotic arm and filter storage closet are attached to the outside of the telescope tube and access the tube interior via existing access ports as shown in Figure~\ref{fig:fe_crosssection}.  The robotic arm and storage closet are co-registered via the robot/telescope bracket, ensuring the position and orientation of the filters are always maintained.  Thus, the robotic arm only needs to be calibrated to the filter storage closet once after installation to the telescope. 

Unlike a more conventional filter changer mechanism which is restricted in motion to one or two dimensions, a robotic arm can move in three dimensions with additional rotational degrees of freedom with little \textit{a priori} restriction. We limit the degrees of freedom of the robotic arm using manufacturer-provided, inherent safety features of the KUKA arm system.  These safety features include active force/torque limiters, speed limiters, and self-diagnostic software monitoring and reporting. Such levels of protection are sufficient for human interaction safety and are in widespread industrial use.

The main subsystems of the ZTF filter exchanger (FE) are the robotic arm, a network of external position sensors, the robot gripper, a filter storage closet, and the set of filter frame assemblies (including filters.)  For the initial ZTF surveys, three optical filters were procured with properties displayed in Figure~\ref{fig:filter_transmission}.

\subsection{KUKA arm}\label{KUKA_arm}
\begin{figure*}
\begin{center}
\includegraphics[scale=0.7]{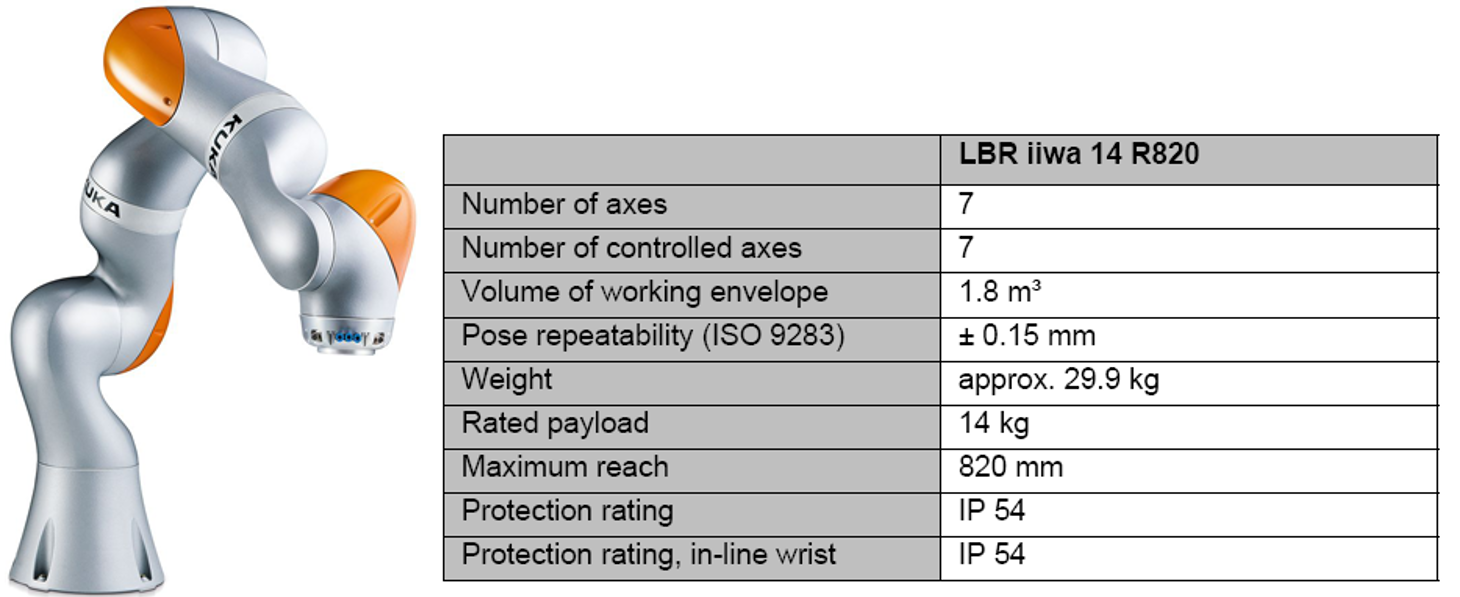}
\caption{Specification of the KUKA LBR iiwa 14 R820 robotic arm.}
\label{fig:fe_LBR_specs}
\end{center}
\end{figure*}

The ZTF robotic arm is manufactured by KUKA Robotics, model LBR iiwa 14 R820. ``LBR'' stands for ``lightweight robot'' and iiwa stands for ``intelligent industrial work assistant''. The ``14 R820'' indicates a 14~kg payload capacity with a maximum reach of 820~mm. An overview of the LBR iiwa 14 R820 specifications is shown in Figure~\ref{fig:fe_LBR_specs}. 

The FE subsystem is also routed through a facility-wide emergency stop (E-stop) system, which can be triggered by the following: \\
1.  The KUKA controller, which deactivates the drive motors and applies the brakes;\\
2.  Panic buttons, placed strategically throughout the dome. These are large, red, latching buttons that can be easily pressed. These allow anyone in the vicinity to respond to an emergency situation that might involve the KUKA arm causing harm to personnel and/or equipment;\\
3.  Door switches, such as telescope tube access doors, filter storage closet access doors, etc. Any entryway or removable panel that would provide human access to the KUKA arm is fitted with a switch. These would prevent accidental human contact with a moving arm.\\

All possible emergency stop inputs listed above are handled by a Banner SC26-2DE Safety Controller. The Banner Safety Controller is a commercial control system which monitors a variety of input devices (buttons, switches, etc.) and uses programmable Boolean logic to determine the output. The Banner Safety Controller output is then used to close the safety signal inputs on the KUKA arm controller to enable operation in a fail-safe architecture.

\subsection{Exchanger sensors and switches}\label{FE_sensors_and_switches}
We use electro-permanent magnets (EPMs) as both latches and sensors in the FE subsystem. Unpowered, these devices exert a strong magnetic force; when powered, the magnetic holding force is released. Driven with a PWM waveform and read with an appropriate circuit, these magnets are used as a sensor that indicates not only the presence a ferrous material but, for proximity less than a couple millimeters, also the distance of that material to the magnet, to an accuracy of better than 50~$\mu$m.

The EPMs each exert a holding force of 50~N at a distance of $\sim$~0.25~mm. The distance to the mating ferrous inserts installed on either side of the filter frame which align with the EPMs are well less than 0.25~mm when the filter is in place, so the total holding force of three EPMs exceeds 150~N. The filter and filter frame together have a mass of 4.25~kg, so that, under gravity, the filter will exert a force of less than 43~N.  This provides a safety margin factor of at least 2.5, making the system robust to individual magnet failures. In the event of EPM failure, the filter is protected by a redundant passive mechanical latch system (Section~\ref{ssec:latches}.)

\subsection{Mechanical latches} \label{ssec:latches}
Mechanical latches are used to physically capture and secure the filter frames at all times, during storage, during hand-off, during transition from storage to the instrument, and while installed on the instrument. These latches are spring-loaded and automatically provide a positive mechanical lock onto the filter frame. Hall effect sensors integral to the latch mechanism indicate the open/close state and can be read remotely. Hall effect sensors are not integral to the operation of the latch and only provide latch state telemetry. Failure of a Hall effect switch neither results in, nor indicates a failure of the latch.

The purpose of the mechanical latches is solely as a safety mechanism to provide mechanical containment of the filter frame. Although capable of safely holding on to the filter frame, the mechanical latches are not used to provide the primary holding force nor as a gripper manipulation device. The EPMs are the primary holding mechanism and the mechanical latches provide backup.

\subsection{KUKA arm gripper}\label{ssec:KUKA_arm_gripper}
The arm gripper is the end-effector attached to the flange at the wrist of the KUKA arm. The gripper has three EPMs that actuate and hold the ferromagnetic targets on the filter assembly.  The EPMs are  magnetized at all times, except when momentarily demagnetized during a hand-off operation, which is done by applying power to the EPMs. Two interlocking mechanical latches accept the headed pins of the filter frame and are only actuated in the presence of the matching set of interlocking latches on the window frame or filter storage closet.  Thus, the filter is always captured by a set of latch jaws and cannot be dropped.  Hall effect sensors provide active telemetry regarding the open-close state of each mechanical latch.  On any anomaly, software watchdogs intervene to halt filter exchanges and stow the telescope in a safe configuration pending human diagnosis.

\subsection{Filter assemblies}\label{ssec:filter_assemblies}
\begin{figure}
\includegraphics[width=\columnwidth]{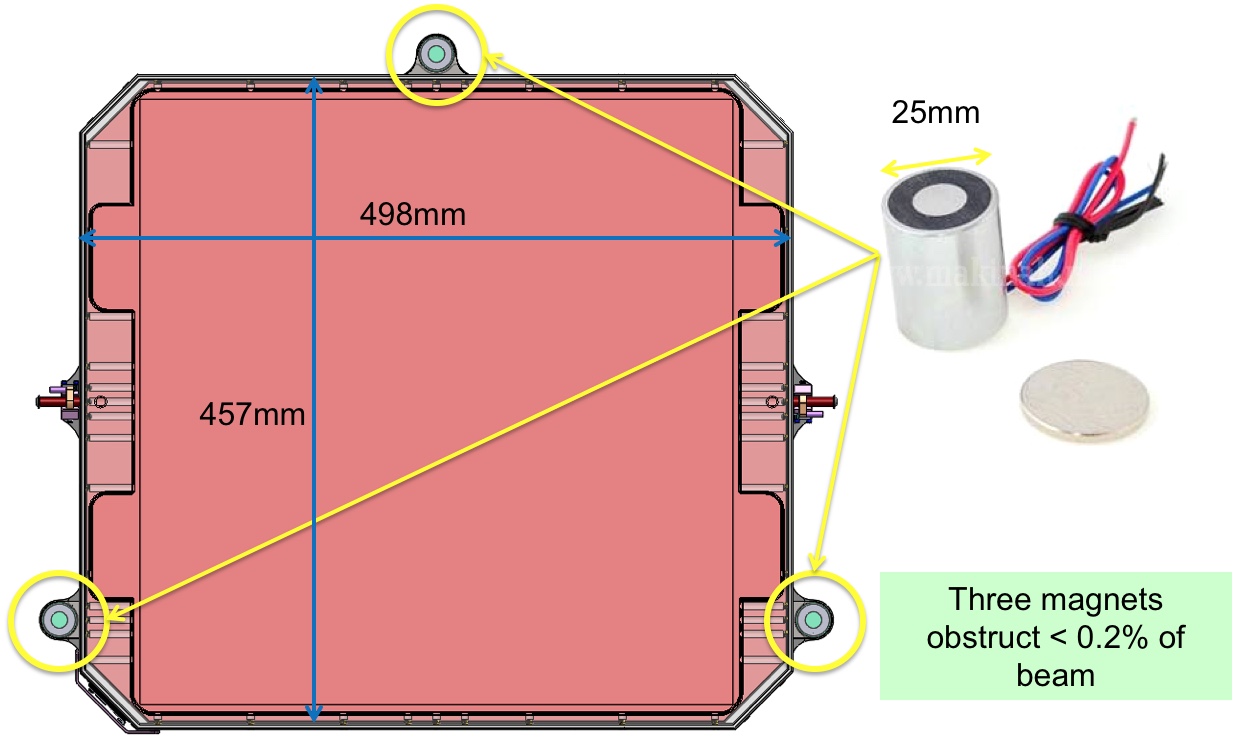}
\caption{One of the ZTF filters, shown within the filter frame that provide interfaces for the safe hand-off off to and from the cryostat window frame.  Pins used for the back-up mechanical latches are seen on the frame exterior, while ferrous (soft iron) strike plates used in the EPM exchange are visible in three locations on the frame perimeter.  The frame boundary is set to allow illumination of the focus CCDs within the cryostat detailed in Section~ \ref{ssec:cryostat_design}.}
\label{fig:filter_frame}
\end{figure}

Each ZTF filter assembly consists of the planar glass spectral filter (Section~\ref{sssec:filters}) and its mechanical frame.  As shown in Figure~\ref{fig:filter_frame}, each frame has three low carbon steel brackets that serve as ferromagnetic targets for the EPMs on the instrument window and robot gripper.  Tapered pins on the instrument-side of the filter frame provide precision location and repeatability of the filter to the instrument via precision reamed holes in instrument window frame.  Two headed pins along the centerline are the interface for the matched pairs of latching jaws on the window and robot gripper.

\subsection{Filter storage closet}\label{ssec:filter_closet}
\begin{figure}
\includegraphics[width=\columnwidth]{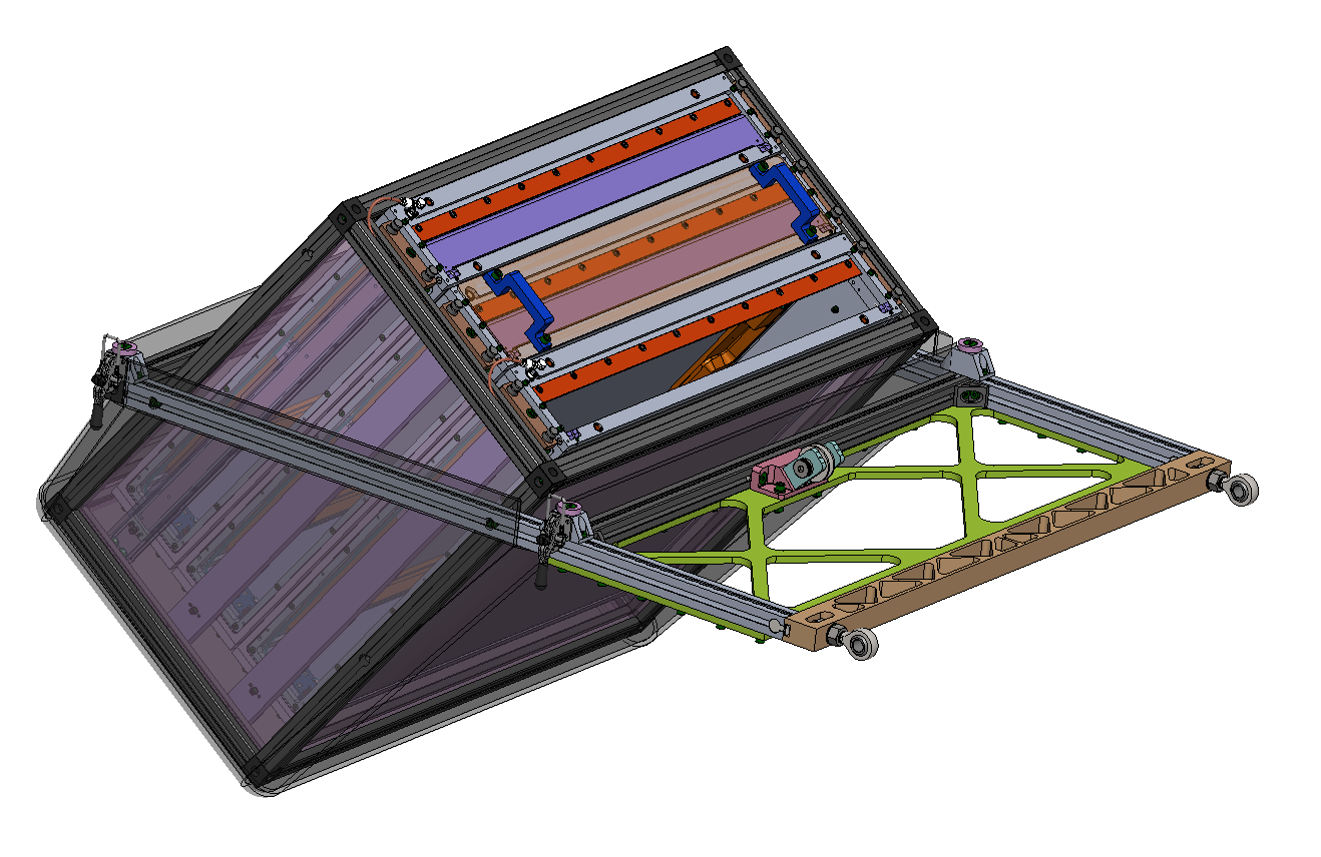}
\caption{The filter storage closet assembly houses and protects three modular filter cartridge assemblies. Each filter cartridge assembly can be interchanged to allow for various survey-specific filter configurations.}
\label{fig:fe_closet}
\end{figure}

The filter storage closet assembly consists of a chassis with removable panels, latching filter storage bays, and robot access doors, as shown in Figure~\ref{fig:fe_closet}.  The latching filter storage bays mimic the instrument window interface and have three EPMs located in the window frame that actuate and hold the ferromagnetic targets on the filter assembly.  Two reamed holes in each storage bay mate with tapered pins on the instrument-side of the filter frame to provide precision location and repeatability of the filter to the storage closet.  

\section{Flat Field Illuminator}\label{sec:illuminator}

Flat fields are  exposures of structureless scenes of  uniform brightness.  They are used to calibrate CCD images for differences in throughput and gain on short spatial scales: between CCDs by comparing intensities across nearby boundaries, and from pixel to pixel within a CCD. Stable color, intensity distribution, and total brightness are prerequisites for achieving high photometric precision.

\begin{figure}[h]
\includegraphics[width=\columnwidth]{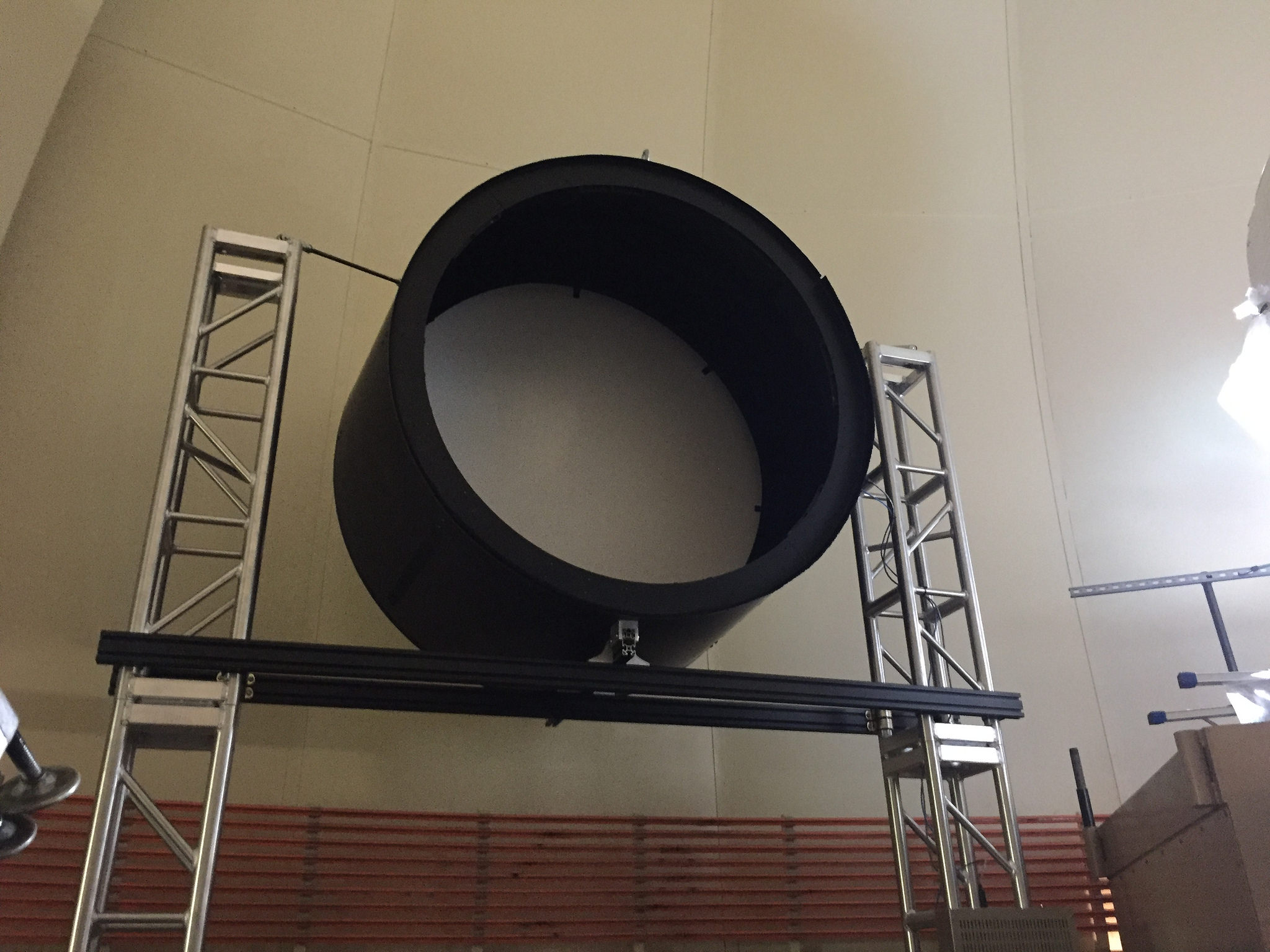}
\caption{The ZTF flat field illuminator installed in the telescope dome.}
\label{fig:ffi}
\end{figure}

To improve the photometric precision of ZTF relative to that achieved with PTF, a new flat-field illuminator (FFI) calibration unit was constructed (Figure~\ref{fig:ffi}).  The FFI consists of a light-tight baffle, into which the telescope can be pointed toward a uniformly illuminated Lambertian screen.  Closely spaced baffles with sharp edges, painted with Avian Black on top surfaces and flocked on the underside, block grazing incidence reflections form the enclosure which seals to the top end of the telescope to greatly attenuate sensitivity to ambient light in the dome to improve the quality of flats acquired in twilight.

\begin{figure*}
    \begin{center}
    \includegraphics[scale=0.5]{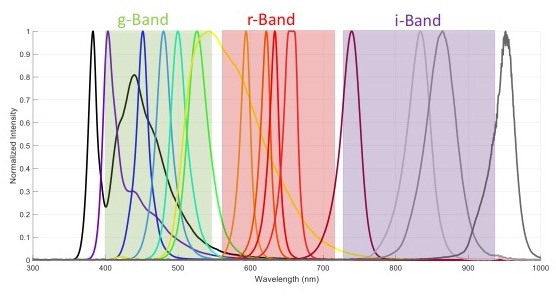}
    \caption{The spectral distribution of ZTF FFI LEDs.}
    \label{fig:ffi_leds}
    \end{center}
\end{figure*}

The illumination source is comprised of eight identical LED boards that encircle the entrance pupil.  Uniformity is enhanced by pointing the LEDs slightly away from the center of the white Lambertian screen. Each  board houses four LEDs in each of 15 unique colors chosen to illuminate the system in each science passband, or out of band illumination can be generated to test for filter leaks.   The spectral response for each LED is plotted in Figure~\ref{fig:ffi_leds}. This distributed LED source was calibrated to provide the entrance pupil with illumination of better than 10\% spatial uniformity for all colors and operating temperatures. The spatial uniformity was confirmed in lab by a collimated photodiode rastered over the entrance pupil and by imaging the screen directly.

While we support a single exposure in which more than one LED is pulsed (sequentially) it is normal practice to acquire multiple (10-12) flats per filter to improve signal to noise and suppress cosmic rays.  It is then preferable to acquire the same number of photons and the same number of exposures but only activate one wavelength at time.  This allows the color balance of the flats to be adjusted post facto.  The improvement in photometric accuracy that this yields is under investigation but appears to be <~1\%.

Another use of the flat field screen is to measure the effect on photometry of the time at which the shadow of the shutter crosses the beam obstructions.  The spatial dependence is a parallax effect due to the fact that the beam obstructions are not in the same plane as the shutter but maps to an effective increase in exposure time, $\Delta(x,y)$, in the sense that for a given exposure time, $T$, signal must be multiplied by $1+\Delta(x,y)/T$ to remove the effect.  The map of $\Delta(x,y)$ by \cite{Giomi:19:ExposureTimeCorrection} shows variation from zero at center of top and bottom edges to 60~ms at left and right edges. This is not to be confused with shutter timing errors which are typically <~2~ms.

\section{Robotic Observing Software}\label{ROS}
The ZTF Robotic Observing Software (ROS) was developed based on the Robo-AO observing system \citep{Riddle:12:SPIE}; the robotic structure and underlying tools were retained and formed the basis of the ZTF robotic system, while new software interfaces were developed for the ZTF hardware.  The ZTF camera system interfaces to five separate computers (one per Archon controller), creating a significant challenge to for ROS manage data processing, FITS header management, and efficient operations.

ROS runs the ZTF Observing System continuously with little to no human involvement, handling nearly all observing conditions autonomously. During the day the system manages data transfer and backup, monitors the weather, and prepares for the observing night.  In the late afternoon, ROS powers on all equipment and completes bias, dark, and flat calibration images (Section \ref{sec:illuminator}); if a problem arises, calibrations are completed in the morning.  ROS then opens the dome and waits for the Sun to pass below $-12$ degrees altitude, then automatically starts science operations. 

Each science observation starts with a request to the queue system to obtain the information for the next exposure.  If necessary, a filter exchange is executed, then the telescope is pointed at the target while the mosaic camera readout process completes, and the hexapod is shifted into position to keep the camera in focus.  Once all observation preparations are complete, the system starts the mosaic observation and data download, while the FITS and data systems execute their operations.  In parallel with all this, the telemetry of each subsystem is monitored for errors, and any errors are handled by the Error Management System (EMS) which may require stopping an observation for more serious problems.  Once the observation has completed, ROS checks that the shutter has closed properly, and then the sequence starts again.  
 
\subsection{ROS computer systems}
The ZTF ROS computers were built using off the shelf parts and assembled in house.  Ten total computers were constructed; one computer is connected to each of the four Archon camera controllers, and a robotic control computer (which also controls the focus CCDs), plus one spare of each computer.  All ten computers are identical; the robotic control computer is able to handle all software operations, and the camera computers are only stressed when reading out the Archon controllers.  Each computer contains three hard drives: the operating system drive (128~GB SSD), a data drive (256~GB SSD), and a large format disk to store data (4~TB hard disk).  

All ROS computers are installed in a 48U mobile rack located in the environmentally controlled Oschin Telescope control room.  A complete set of spare ROS computers are available, allowing for immediate swapping in case of computer failure and continued operation of ZTF beyond the initial survey period.  If a computer faults, a cold swap can be made by moving just three cables (Ethernet, fiber, then power) from a primary to a spare, each of which is annually tested as ready to run upon upload and recompilation of the latest released ROS version.  

\subsection{Software architecture}
ROS is based upon a modular, fail-safe, multi-threaded, multi-daemon software architecture coded in C++, where each daemon controls a major subsystem (which can consist of multiple hardware elements).  A supervisor daemon manages all operations of every element of the instrument; it determines the course of action during an observation and manages any errors that occur.  There are thirty total daemons running the ZTF OS.  ROS has been designed to be able to run continuously for an extended period, while allowing staff to monitor the system to determine its operational performance, track nightly errors, and reconfigure parameters if necessary.  

ROS operations are managed through the use of configuration files, which support both engineering and science operations, striking a balance between exposure of control parameters to users and self-checking systems to maintain system safety and efficiency.  Manual control of the hardware is executed through two separate programs, one that allows limited control of the robotic observing system itself (in order to do simple things like start, stop, or reset subsystems), and a second that allows full interface to all robotic functions under manual control.

A watchdog system has been implemented on all five of the ZTF computers.  On the camera computers, the watchdogs only make sure that each is running, while on the control computer it is also tasked with restarting the robotic control daemon if it fails.  A planned upgrade to the EMS will use the watchdogs to detect if an entire computer system has shut down and reconfigure ZTF to continue observing (this requires manual intervention at present).

\subsection{Operations monitoring}

ROS produces extensive, detailed logs of all operations throughout the day, including a separate error log of all errors produced by the observing system; around one million log entries are created per day.  A similar amount of telemetry is created by all subsystems; this telemetry tracks the operation of the subsystem hardware and software, with entries for physical values (temperatures, positions, etc.) and software states.  Examining these information sources provides insight into every aspect of ROS, but is also overwhelming due to the amount produced.

To alleviate this, a website is used to monitor the system as it operates.  This website is color coded to flag when an error occurs, and includes plots for sensor data and an archive of all log and telemetry created by the system.  ROS also sends out alert emails when a major error occurs (such as a failure to finish a filter exchange or start the night properly); these emails are repeated every six hours if not cleared by human intervention.  Warning messages for less-critical events are also sent out, and ROS also sends a morning update message with details about the performance of the observing system during the previous night. Other software utilities are used for mining the data logs for specific information; for example, timing the filter exchange process for monitoring performance, or determining failure rate statistics.

A separate, science-focused internal operations website is available for monitoring ZTF operations, assessing delivered image quality and open-shutter efficiency, tracking ROS logs, and curating completed observations.  This site provides data quality metrics provided by the ZTF data system image processing pipeline running at IPAC.  The operations team has access to basic data products and data quality metrics for each image approximately 8-12~minutes after the close of the ZTF exposure shutter, depending on network traffic and field stellar density.

\subsection{Mosaic camera control}

The mosaic camera subsystem consists of five control daemons, one per computer; four of these are identical and handle the science CCD operations, while the fifth controls the focus CCDs.  Commands are sent to the five Archons in parallel, and all five systems execute the same commands.  An additional set of commands is used to handle focus operations which calculate the hexapod positions to keep the entire mosaic in focus.

\subsection{Filter exchanger control}

The filter exchange system underwent extensive testing of the hardware components to understand their operation and failure modes; this included testing the entire FE system in a cold box as well as cycle testing on the Oschin telescope to ensure reliability.  Software interfaces for the hardware are unified under a filter exchange daemon, which manages all operations of the filter system.  This daemon is able to manage the operation of the exchanger elements, and automatically recover from most of the errors that may occur during an exchange.

ROS oversees the operation of the exchanger; an exchange requires moving the telescope to a defined position because the KUKA arm cannot currently operate under arbitrary gravity vectors.  ROS also manages the errors that the filter daemon cannot, such as power cycling the hardware in order to reset in cases where it has become unresponsive.  If the filter system is unable to automatically recover from a fault, ROS will automatically reconfigure to use only the currently installed filter and continue science operations.

\subsection{FITS header synchronization}

ZTF has the challenge of synchronizing FITS header information across five separate computer systems.  This is handled through the use of a separate subsystem daemon that is solely responsible for gathering this information and distributing it to all systems.  The daemon is the same code across all five computers, with the daemon on the robotic control computer as the master, while the daemons on the camera computers are all nodes.  Each of the subsystem daemons creates a FITS header stub file which contains the information that particular daemon should include in the FITS header, which are then combined by ROS into the full header.  

At the start of an exposure, ROS signals the FITS subsystem to clear out all old data and gather header information from each subsystem.  All of the header parameters known at this point are then shared with the FITS nodes, so each computer has the same header information.  At the end of the exposure, the last remaining items required for the FITS header are then gathered and synchronized across the system.  At this time, parameters that require calculations can also be created and shared, as well as any saved parameters inside the robotic system.  

Once all of this information is synchronized, ROS signals the FITS system to create a master FITS header file that is read by the FITS creation software when the science CCD images are created.  The CCD camera control system contains the local FITS header information (i.e. camera name, CCD identifiers, image size, etc) which it writes directly into the FITS header.  The synchronized FITS header information is completely cleared from the system before the next CCD exposure is started to ensure that header information is kept unique for each CCD exposure.

\subsection{Data management}

After the science image FITS files are written to the data drive, they are compressed using a lossless Rician algorithm through the use of fpack, a part of the CFITSIO package \citep{CFITSIO}.  ROS commands the data subsystem to do this compression, and then download the data to the IPAC servers in Pasadena through the High Performance Wireless Research \& Education Network (HPWREN), a rural microwave link data network operated by the University of California, San Diego in collaboration with the San Diego Supercomputing Center.  The original images for each CCD are compressed down to about 24~MB total (a factor of roughly 6), so each mosaic image is in total just under 400~MB compressed.  Each camera computer then automatically sends the data in parallel to IPAC; it takes around 20~s to send the entire mosaic image to IPAC, so the mosaic is normally received at IPAC 40-50~s after the shutter closes.  

If the network connection is slow or fails, the data management system is able to continue to support observations while attempting to download data.  New images are added to the download list, and if they fail they go to the end of the list and are tried again; this allows new images to keep downloading in near-real-time while catching up with older data as fast as possible.  The system will continue to attempt to download data through the next day if necessary, until it has to prepare for the next night of observing.

Each night that the ZTF OS operates, it produces up to 250~GB of compressed FITS files, and each day these files have to be moved to the backup drive, otherwise the data drive will fill up and the OS will no longer be able to produce data.  To transfer data, ROS automatically moves the data to the backup drive and clears the data drive; if the backup drive is too full of data then ROS will remove old data until there is enough space for the previous night's data.  ROS retains about one month of data on the system that can be down-linked if necessary.

ROS integrates all of the above elements, and many more (weather and sensor monitoring systems, shutter control, and power management among them) into a fully automated observing system (Section~\ref{ssec:efficiency}).  

\section{Observing System Performance}\label{performance}
\subsection{Image quality error budget}\label{DIQ_error_budget}

ZTF is designed to detect optical transients in a brightness range that may be spectroscopically investigated using available larger telescopes.  As such, we prioritize wide-field of view with modest image quality.  Based on experience with image differencing and false positive rejection, we selected 1-arcsecond per pixel sampling and set the optical design challenge to maximize field of view with a bottom-line, delivered image quality goal of 2 arcseconds FWHM in ZTF-r filter and 2.2~arcseconds in ZTF-g filter.  Although not all of our error sources are independent, we used an independence approximation to allocate contributions to the FWHM as shown in Table~\ref{tab:DIQ_budget}.  This error budget was used extensively to flow optical performance requirements into mechanical tolerances for the telescope, cryostat, and filter exchanger subsystems.

\begin{table*}
	\begin{center}
	\caption{DIQ error budget for ZTF in r-band, FWHM in arcseconds\label{tab:DIQ_bud}}
	\label{tab:DIQ_budget}
	\begin{tabular}{l l l l}
	\hline
	\multicolumn{3}{c}{Atmosphere at Zenith Angle = 30~degrees} \\
	\hline
Median Free Atmospheric Seeing & 1.10 \\
Dome and Mirror Seeing & 0.35 \\
Atmospheric Refraction (ZTF-r band) & 0.07 \\
& & 1.16 \\
	\hline
	\multicolumn{3}{c}{Telescope} \\
	\hline
M1 Figure & 0.40 \\
Tracking Errors* & 0.50 \\
Vibration & 0.39 \\
Hub Tilt Rel to Optical Axis & 0.40 \\
M1 (Optical Axis) Tilt Rel to Cell & 0.39 \\
Schmidt Plate Axial Position & 0.04 \\
Schmidt Plate Decenter & 0.10 \\
Schmidt Plate Tilt & 0.10 \\
Schmidt Plate Aspheric Coeff & 0.19 \\
Schmidt Plate Index of Refraction & 0.28 \\
Schmidt Plate Abbe Number & 0.21 \\
& & 1.02 \\
	\hline
	\multicolumn{3}{c}{Instrument} \\
	\hline
Optical Design IQ (full field avg) & 0.83 \\
Cryostat Decenter & 0.16 \\
Deviation from Best Focus (Hub motion) & 0.31 \\
Cryostat Window Rel to FPA & 0.03 \\
Cryostat Window Tilt & 0.00 \\
Cryostat Window Opt v. Mech Axis & 0.05 \\
Cryostat Window Center Thickness & 0.07 \\
Cryostat Window Glass Melt Index, n	& 0.08 \\
Cryostat Window Thermal Variation in n & 0.06 \\
Optics Manufacturing Surface Errors & 0.23 \\
Mosaic Tilt Rel to Cryostat	& 0.21 \\
CCD Surface Relative to Plate & 0.27 \\
FPA Plate Height Relative to Hub & 0.11 \\
Field Flattener Tilt & 0.05 \\
Field Flattener Decenter & 0.10  \\
Field Flattener Opt v. Mech Axis & 0.16  \\
Field Flattener Power & 0.10 \\
Field Flattener Final Temperature & 0.02 \\
CCD Lateral Diffusion & 0.48 \\
& & 1.02 \\
	\hline
Total DIQ (FWHM, arcseconds) & & & 2.00 \\
	\hline
	\end{tabular}
	\end{center}
\end{table*}

\subsection{Delivered image quality}\label{ssec:DIQ}By Fall 2018, the ZTF Observing System was fully operational and generating significant numbers of transient alerts in real-time for the US astronomical community every clear observing night, while DIQ was continued to be improved through automation of focus sensing and hexapod control.

\begin{figure*}
    \begin{center}
    \includegraphics[scale=0.55]{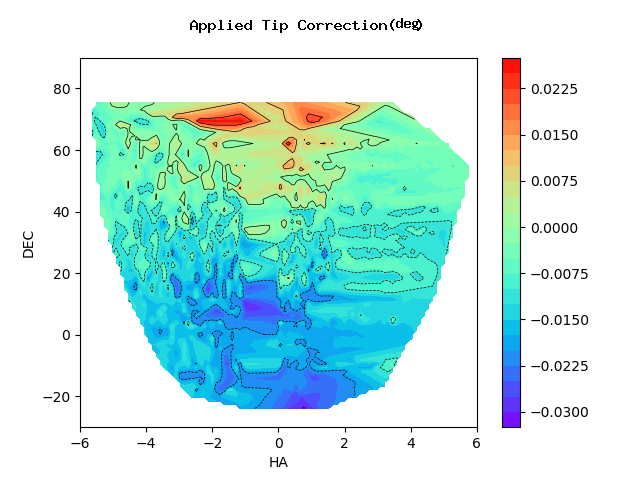}
     \includegraphics[scale=0.55]{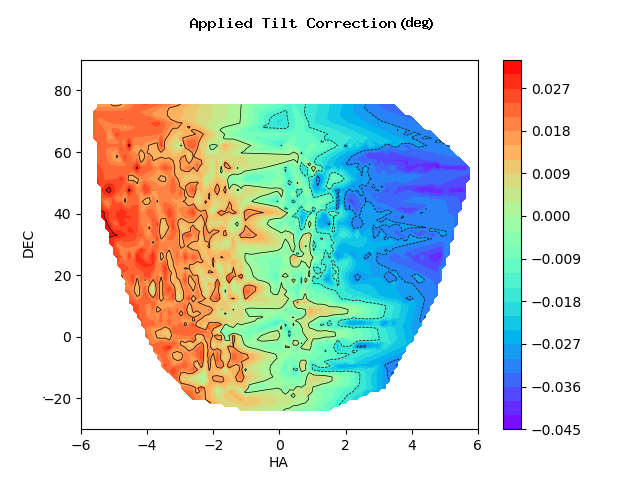}
    \caption{Open loop corrections for tip and tilt, in degrees as a function of Hour Angle (hours) and Declination (degrees).  These maps are derived represent hexapod settings for sharpest images at each pointing and are being steadily refined.}
    \label{fig:TipCorrection}
    \end{center}
\end{figure*}

\begin{figure*}
    \begin{center}
    \includegraphics[scale=0.7]{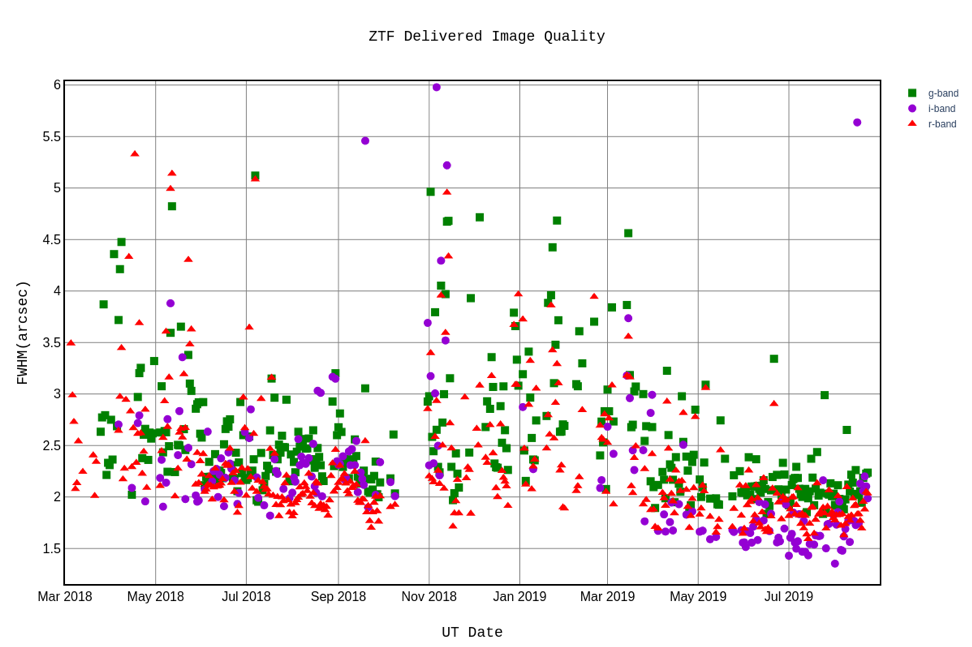}
    \caption{Evolution of  median of FWHM over full field and all exposures for any given night in g-band (squares), r-band (triangles) and I band (circles).  The majority of the time seeing is subdominant and differences are largely due to optical aberrations.  At the 1”/pixel image scale PSF sampling is routinely sub-nyquist in r and i bands.}
    \label{fig:DIQ_by_date}
    \end{center}
\end{figure*}

As expected, flexure in the instrument supports causes significant tip and tilt, and a small amount of piston.  The hexapod is driven to compensate at every new pointing based on empirically determined flexure maps derived for each pointing by averaging hexapod settings when good image quality is achieved (Figure \ref{fig:TipCorrection}).   Initially full mosaic tip, tilt and focus maps were made by extra focal imaging during boustrophedonic (raster) scanning of the entire sky. However, it was found that the required correction also depends significantly on the magnitude and  direction of the telescope slews, which were artificially small in the case of the reversing raster scans.  Therefore the open-loop correction of tip and tilt is supplemented by a servo control loop which removes the effects of both mechanical hysteresis and defocus due to thermal expansion.  Predetermined focus offsets are also applied during filter changes to compensation for difference in optical path length between filters.  Full recovery from a focus transient currently takes 4 to 5 exposures, so, for large slews, DIQ is still heavily dependent on the quality of the open loop correction.  A separate auto-guider was deployed on the bore-sighted finder telescope allowing extra-focal imaging at all four locations around the perimeter to improve SNR on focus measurements with one of the four positions being on the opposite of focus.   

Focus sensing was significantly complicated by edge-of-field vignetting effects in our off-axis CCDs.  While telescope vignetting was expected, we did not expect the much steeper shadowing effects that were introduced when our ZTF-r and ZTF-g science bandpass filters were erroneously specified with smaller optical area than desired, leading to field-dependent chromatic effects in the focuser subsystem.  Filter coatings cover all beams reaching the science mosaic, but only partially cover the beams reaching the focuser detectors. This error was caught and avoided when specifying our ZTF-i filter.  In response, we improved the robustness of our out-of-focus donut extraction algorithm to properly handle field-dependent vignetting by subdividing the field of view of each focuser into 8x8 calibration sub-regions allowing mapping of the field dependence. 

As the flexure compensation system systematically addressed these issues, a marked improvement in temporal stability and DIQ uniformity across the field was achieved Figure~\ref{fig:DIQ_by_date},  with all control features became concurrently operative by April 2019:  better open loop offset maps, focus offsets between filters, servo control based on four extra focal image sensors with field dependence corrected, and a separate autoguider.   

\begin{figure*}
    \begin{center}
     \includegraphics[scale=0.2]{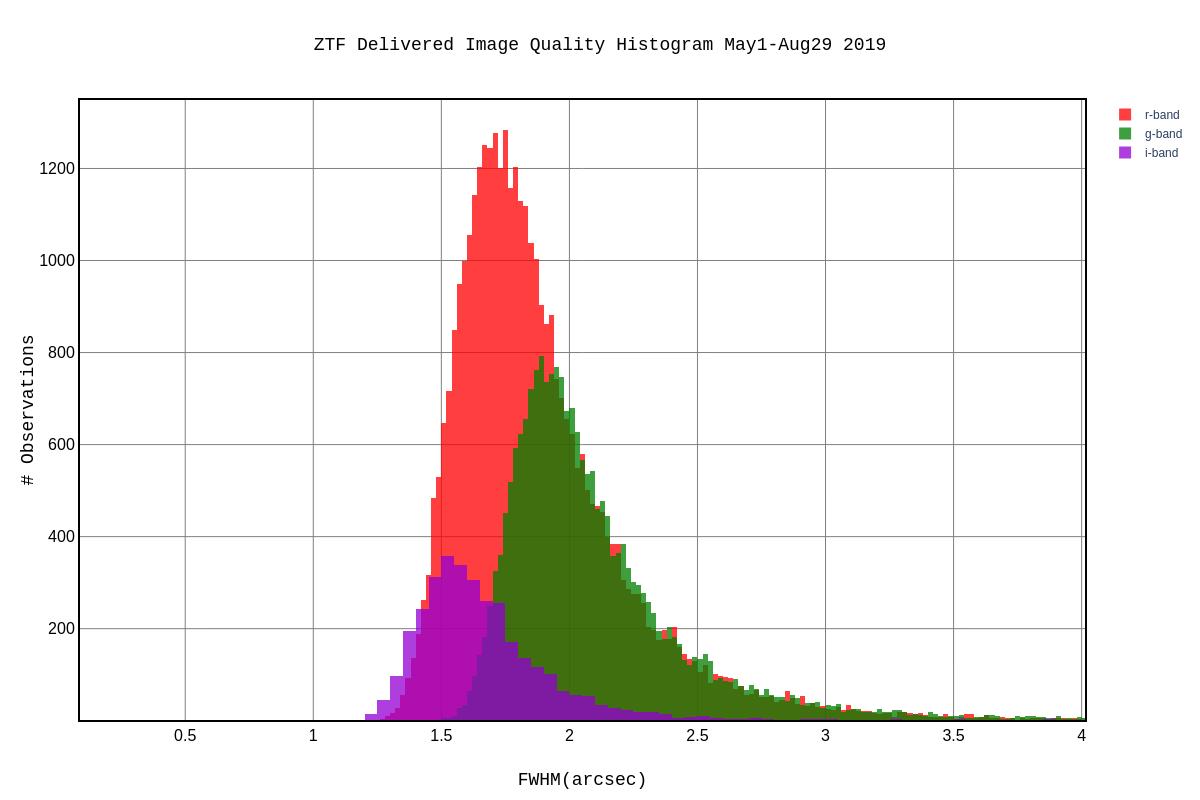}
    \includegraphics[scale=0.29]{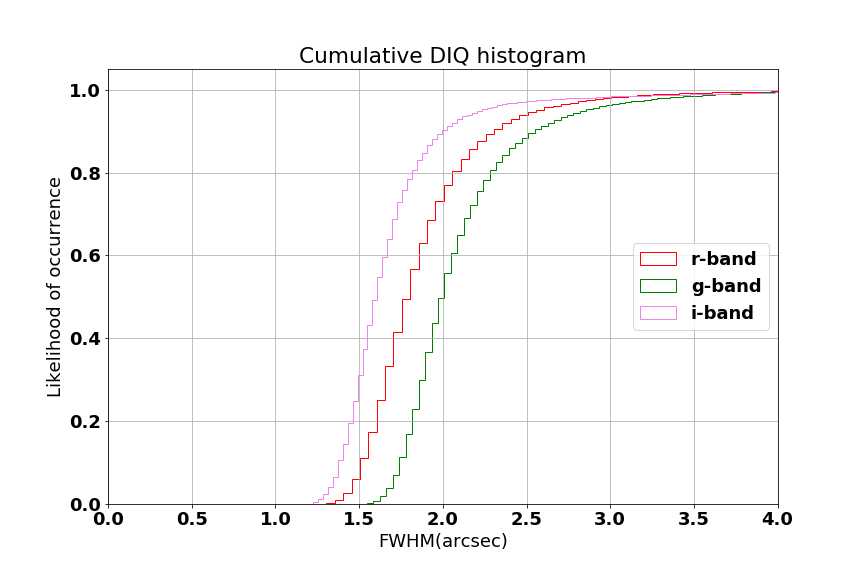}    
    \caption{Histograms of DIQ in g, r and i bands, measured across the entire field of view for all images from May 1 through August 29, 2019. All weather, airmass and seeing conditions have been included.}
    \label{fig:cumulative_DIQ}
    \end{center}
\end{figure*}

Residual design aberrations dominate seeing in all bands until seeing exceeds ~2 arcsec with chromatic errors being slightly greater in ZTF-g band as models predicted. Figure~\ref{fig:cumulative_DIQ} shows cumulative distribution of DIQ including all field points, all air mass and seeing conditions and exposure times. Median DIQ is better than the budget allocation in all bands, and though our specification was for median conditions, we meet target DIQ 2/3 of the time.)  

\begin{figure*}[htp]
    \begin{center}
    \includegraphics[scale=0.4]{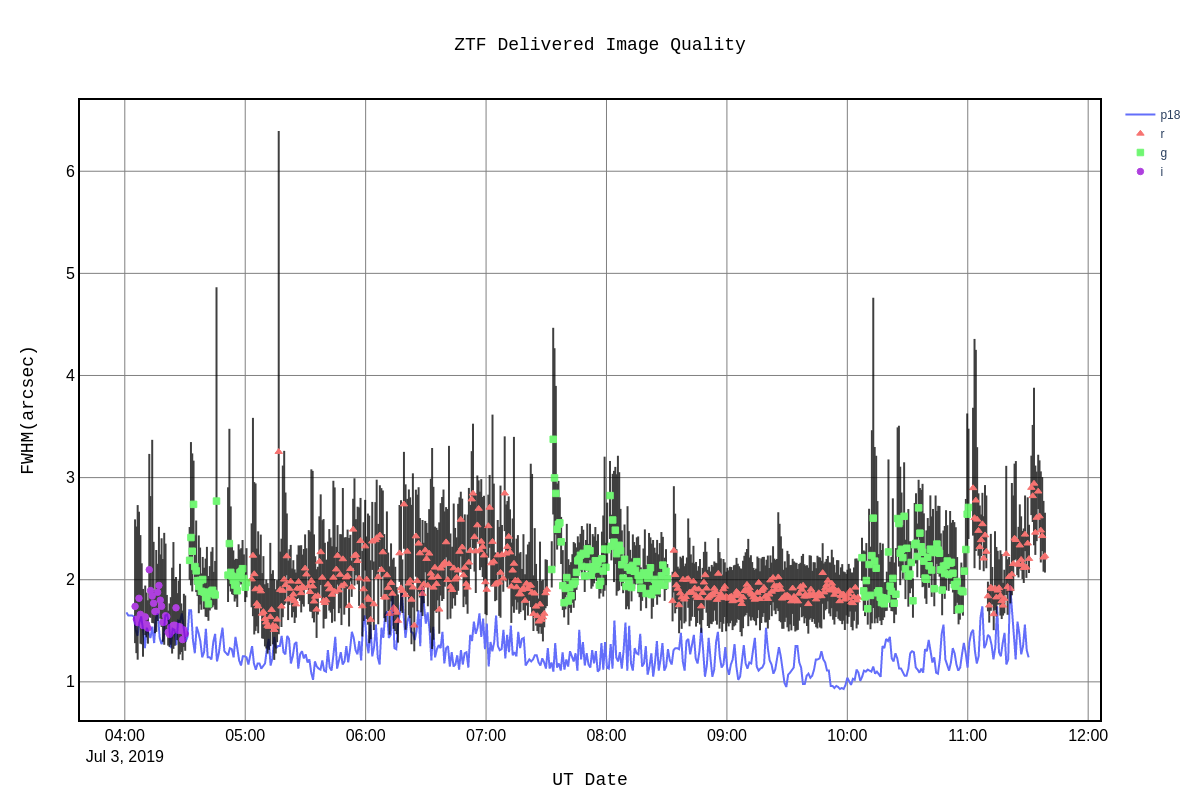}
    \caption{Median FWHM for each frame for the night of July 3, 2019.  g, r, and I bands are marked by squares, triangles and  circles respectively, while vertical lines indicate range across a single field.   The observations in r-band from 08:35 to 10:10 are more stable and exhibit less tilt since the same field was observed repeatedly.  Compare to the period from 05:05 to 07:35, where pointing changes introduce disturbances that are not fully corrected by the static flexure map.   DIQ at low to moderate airmass is  uncorrelated with seeing as reported by the P18 Differential Image Motion Monitor (blue line) unless seeing exceeds 2 arcsec.}
    \label{fig:DIQ_20190703}
    \end{center}
\end{figure*}

The temporal variability in performance is apparent in the time series of DIQ shown in Figure~\ref{fig:DIQ_20190703}.  This example tracks both full-field median (dots) and individual CCD extrema (black bars) DIQ values.  During a period of several hours (around UT 9:00) a single observing field was observed continuously, allowing us to estimate the precision of the servo measurement.  Outside of this single-field period, the DIQ stability is degraded by hysteretic residual flexure, which depends on the slew trajectory history. The potential for flexure, apparent in wide excursions between best and worse detector DIQ motivates on-going efforts to improve both open loop map accuracy and servo settling time. 

\subsection{Observing system efficiency}\label{ssec:efficiency}

\begin{table*}
	\begin{center}
	\caption{Observing sequence median execution times (in seconds) for 503 observations on a fixed target.  All times are independently measured from the ROS control computer system clock with time tags inserted into the software at key points in the sequence.}
	\begin{tabular}{l l l l}
	\hline
Mosaic readout arm & 0.009 $\pm$ 0.018\\
Mosaic readout & 8.32 $\pm$ 0.11 \\
Mosaic preparation & 0.09 $\pm$ 0.04 \\
Observation preparation (total) & & 8.44 $\pm$ 0.11 \\
	\hline
Mosaic arm & 0.05 $\pm$ 0.07 \\
Science exposure & 30.008 $\pm$ 0.034 \\
Shutter closure & 0.450 (fixed time) \\
Science observation (total) & & 30.50 $\pm$ 0.08 \\
	\hline
Total observation sequence time & & 38.98 $\pm$ 0.14 \\
	\hline
	\end{tabular}
	\end{center}
	\label{tab:obs_efficiency}
\end{table*}

The ZTF requirement for observing overhead is 15~s, with a 10~s goal, for adjacent fields (separated by 7.5\degree), when not limited by dome motion.  Meeting the requirement overhead was only possible by overlapping the mosaic exposure and FITS file download and write processes within ROS, as all mosaic operations alone require about 18~s to complete.  Table~\ref{tab:obs_efficiency} shows the median results for a test run of 503 science observations on a single target (to avoid telescope and dome slew effects).  Note that a fixed shutter closure time is used within the observing sequence; as a direct closure time is not measured during observing ROS uses the measured 430~ms close time plus 20~ms padding for shutter closure.  The median time for these observations is 38.98~s; with a 30~s exposure time this is an overhead that is 6~s lower than the ZTF requirement and 1~s lower than the goal time.

The ROS software alone, after taking into account hardware-only delays by examining the observing logs, transactions with the queue scheduler, and measured values from  Table~\ref{tab:obs_efficiency}, requires 40-60~ms per observing sequence,  depending somewhat upon the specifics of each science observation.  For the series of 503 observations in Table~\ref{tab:obs_efficiency}, without telescope slew, the ROS latency timing jitter had a RMS of just 1.4~ms. 

During nightly science operations, the telescope and dome slew times affect the observing overhead.  If the telescope moves between adjacent fields, the telescope and dome slews complete within the observation preparation times discussed above.  However, with filter exchanges (usually about 5-10 per night) and longer slews driven by different science programs, the overhead time increases.  Each filter exchange adds about 60-70~s to a science observation, and the longer slews involved for the exchange increase the time between observations by an average of 40~s while the ZTF observing system waits for the telescope and dome to align on the science target.  Error recovery by ROS during the night also adds to overhead, but fortunately the system has shown almost no errors on an average night that require intervention.  The additional operational overheads tend to average to 2-3 more seconds per exposure over the course of a night; that still results in a lower overhead than the 15~s ZTF requirement for adjacent fields.  Despite observing 30~s per field and moving between observations, ZTF open shutter time is usually above 70\% on any given night, resulting in observation rates of 85 science targets per hour or more.  ZTF has routinely demonstrated 1000 science exposures covering over 47,000 square degrees of sky per observing night; on the shortest night of 2019 ZTF observed 650 science fields in about 7.5 hours.  

\acknowledgments
\section{Acknowledgments}
Based on observations obtained with the Samuel Oschin Telescope 48-inch and the 60-inch Telescope at the Palomar Observatory as part of the Zwicky Transient Facility project. ZTF is supported by the National Science Foundation under Grant No.\,AST-1440341 and a collaboration including Caltech, IPAC, the Weizmann Institute for Science, the Oskar Klein Center at Stockholm University, the University of Maryland, the University of Washington (UW), Deutsches Elektronen-Synchrotron and Humboldt University, Los Alamos National Laboratories, the TANGO Consortium of Taiwan, the University of Wisconsin at Milwaukee, and Lawrence Berkeley National Laboratories. Operations are conducted by Caltech Optical Observatories (COO), IPAC, and UW.

\newpage
\bibliography{references}


\end{document}